\documentclass[aps,pra,twocolumn]{revtex4}
\usepackage{amsmath}
\usepackage{amscd}
\usepackage{graphicx}
\usepackage{subfigure}
\usepackage{appendix}
\usepackage{amsmath}
\usepackage{verbatim}

\begin{document}
\title[Short Title]{Shortcut scheme for one-step implementation of a three-qubit nonadiabatic holonomic gate}

\author{Bi-Hua Huang$^{1,2}$}
\author{Yi-Hao Kang$^{1,2}$}
\author{Zhi-Cheng Shi$^{1,2}$}
\author{Jie Song$^{3}$}
\author{Yan Xia$^{1,2,}$\footnote{E-mail: xia-208@163.com}}

\affiliation{$^{1}$Department of Physics, Fuzhou University, Fuzhou 350116, China\\
             $^{2}$Fujian Key Laboratory of Quantum Information and Quantum Optics (Fuzhou University), Fuzhou 350116, China\\
             $^{3}$Department of Physics, Harbin Institute of Technology, Harbin 150001, China}

\begin{abstract}
Nonadiabatic holonomic quantum computation has attracted increasing
interest because of its robustness. In this paper, based on the
shortcuts to adiabaticity (STA), we propose a scheme to construct a
three-qubit nonadiabatic holonomic gate in a cavity quantum
electrodynamics system. In the scheme, the three-qubit holonomic
gate is implemented in a single step. Besides, numerical simulations
show that the constructed holonomic quantum gate holds the
characteristics of high-fidelity, decoherence-resistance and
error-immunity.
\end{abstract}


\maketitle

\section{Introduction}

 The key step for implementing effective quantum computation is to construct
 robust quantum gates \cite{MontangeroPRL0799}. Recently, researchers have shown
that one promising approach for realizing the resilient quantum gate
is holonomic quantum computation (HQC)
\cite{ZhouPRL17119,ArroyoNC1505,OreshkovPRL09102,JrNAT13496}.
Generally, in HQC, the information is encoded in a degenerated eigenspace of the parameter-dependent Hamiltonian, then by driving the system along an adiabatic loop in the parameter space, a holonomic gate can be constructed \cite{ZhouPRL17119,ArroyoNC1505,OreshkovPRL09102,JrNAT13496}. Since HQC is based on non-Abelian geometric phases
\cite{BerryPSRA84392,AharonovPRL8758,WilczekPRL8452}, it is robust against certain types of control errors
\cite{OreshkovPRL09102,LongPRL13110}. Applications of
adiabatic holonomic gates have also been presented with quantum dots
\cite{SolinasPRB0367}, superconducting qubits \cite{FaoroPRL0390},
and trapped ions \cite{DuanSCI01292}.

Although the robustness merit of the adiabatic HQC \cite{FlorioPRA0673,LupoPRA0776,ChiaraPRL0391,LuopPScr0979}, the adiabaticity brings an unfavorable factor: the time scale for the adiabatic evolution is long which may make the system vulnerable to decoherence. To overcome the disadvantage, quantum computation based  on nonadiabatic and non-Abelian geometric phase, i.e., nonadiabatic HQC (NHQC), has been proposed \cite{SjoqvistNJP1214,XuPRL12109}. Due to its robustness against control errors and high-speed, NHQC quickly attracts much attention \cite{ZhengPRA1693,JingPRA1795,LongPRL13110,ZuNAT14514,JohanssonPRA1286,AbdumalikovNAT13496}. Consider an $M$-dimensional quantum system governed by Hamiltonian $H(t)$ and an $N$-dimensional subspace $\mathfrak{S(t)}$ spanned by a set of orthonormal basis states $\{|\chi_l(t)\rangle~(l=1,...,N)\}$ with $|\chi_l(t)\rangle$ satisfying $i|\dot{\chi}_l(t)\rangle=H(t)|\chi_l(t)\rangle$.
Researchers \cite{SjoqvistNJP1214,LongPRL13110,XuPRL12109} point out, when the system is initially in subspace $\mathfrak{S(0)}$, after the evolution time $T$, the evolution operator $U(T)$ of the system would be a holonomic matrix acting on $\mathfrak{S(0)}$ if the following two conditions are satisfied:
 (i) $\sum_{l=1}^{N}|\chi_l(T)\rangle\langle\chi_l(T)|=\sum_{l=1}^{N}|\chi_l(0)\rangle\langle\chi_l(0)|$;
(ii) $\langle\chi_l(t)|H(t)|\chi_{l^{\prime}}(t)\rangle=0~ (l,l^{\prime}=1,...,N)$. That is, as long as conditions (i) and (ii) are satisfied, quantum holonomy can also arise for a nonadiabatic evolution.

In recent years, to fulfill the conditions (i-ii), a variety of schemes for realizing fast and error-resistant quantum computation has been raised \cite{XuePRA1592,XuSR1404,XuPRA1795,XuPRA1592,TongPRA1795,LiangPRA1489,ZhouOE1523,XuePRA1694,XuePRAP1707,SongNJP1618,LiuPRA1795,ZhangSR1505}. Among these scenarios, we notice that schemes \cite{SongNJP1618,LiuPRA1795,ZhangSR1505} using the technique of
``shortcuts to adiabaticity'' (STA) \cite{VitanovRMP1789,MugaAAMOP1362,Demirplak0308,ChenPRL10105,CampoSR1202,BerryJPA0942,MugaJPB0942,MasudaPRA1184,ChenPRA1183}
are attractive. In STA, researchers design time-dependent Hamiltonian to drive the system evolve along a nonadiabatic path from a given initial state to a prescribed final state in a fast and robust manner. Asides from the application in quantum computation \cite{SongNJP1618,LiuPRA1795,ZhangSR1505}, the STA technique is
also applied in fields like fast entanglement generation \cite{KangPRA1694,WuSR1707}, fast population transfer \cite{ChenYHPRA1489,HuangPRA1796,ZhouNP1713},
and others \cite{TorosovPRA1489,VacantiNJP1416,SalaPRA1694,DeffnerNJP1618,SongPRA1795,OpatrnyNJP1416,CampoPRA1490,BachmannPRL17119}. In addition, some experimental applications for STA have been reported \cite{DuNC1607,AnNC1607,BasonNP1208}.
Usually, the techniques of designing shortcuts
include transitionless quantum driving \cite{BerryJPA0942,Demirplak0308,ChenPRL10105}, ``fast-forward'' scaling \cite{MasudaPRA1184}, and inverse engineering based on Lewis-Riesenfeld invariants \cite{ChenPRA1183,LewisJMP6910}.
Recently, other methods \cite{IbanezPRL12109,IbanezPRL1387,AdelPRL13111,ChenYHPRA1795,BaksicPRL16116} have also been put
forward to improve or extend the STA. For example, Baksic \textit{et al}. \cite{BaksicPRL16116} have proposed to significantly speed up adiabatic state transfers by using dressed states. Chen \textit{et al}. \cite{ChenYHPRA1795} have raised an easy to get Hamiltonian to accelerate the evolution of the system.

On the other hand, although it is known that any quantum gate operation in quantum computation
can be performed only with a sequence of one- and two-qubit unitary gates
\cite{DiVincenzoPRA9551,BarencoPRL9574}, this generic construction
usually becomes complicated
when the number of qubits increases.
Therefore, single-step multiqubit quantum gates seem attractive in practice. In this respect, we note that currently many researches focus on one- and two-qubit holonomic gates \cite{SongNJP1618,XuePRA1592,XuSR1404,ZhouOE1523,XuPRA1795,LiangPRA1489}, single-step $N$-qubit ($N\geq3$) holonomic gate is still a relatively new issue deserved investigation.
In view of this, and inspired by the concepts of NHQC and STA, we develop a scheme to implement a one-step three-qubit nonadiabatic holonomic quantum gate in a cavity QED system. In the scheme,
firstly, we utilize the shortcut technique of using dressed states \cite{BaksicPRL16116} to inversely design a shortcut Hamiltonian. Secondly, based on the designed shortcut, a three-qubit nonadiabatic holonomic quantum gate is presented in a single step. Then, to investigate the practical feasibility of the proposed quantum gate, numerical simulations are performed. The results indicate that the proposed holonomic gate has a high fidelity and is robust against the decoherences and the fluctuations of the control parameters.

The paper is organized as follows. In Sec.~\ref{section:II}, we
introduce the constructing method for inversely engineering the time-dependent Hamiltonian of a quantum system. In part A of Sec.~\ref{section:III}, we scrutinize the physical model for implementing a three-qubit nonadiabatic holonomic gate and give the effective Hamiltonian of the system. In part B of Sec.~\ref{section:III}, we detail the construction of the target nonadiabatic holonomic gate via STA.
In Sec.~\ref{section:IV}, as an illustration, we numerically investigate the performance of the three-qubit Toffoli gate scheme in the presence of the control parameters'
variations and decoherence. Conclusions
are given in Sec.~\ref{section:V}.

\section{INVERSE HAMILTONIAN ENGINEERING VIA DRESSED STATES SCHEME}\label{section:II}

In this section, for the sake of clarity, we briefly review the shortcut technique of using dressed states (for brief, we call it dressed-state-based shortcuts and abbreviate it as DSBS) \cite{BaksicPRL16116}, and apply the technique to inversely design the Hamiltonian of a
three-level system.

Consider the original Hamiltonian of a quantum system is $H_0(t)$, the goal is driving the system evolve from some initial state $|\psi_i\rangle$ to the target state $|\psi_f\rangle$. The adiabatic passage achieves this goal by imposing adiabatic conditions so that the system can evolve along a certain instantaneous state of $H_0(t)$. To free the evolution from adiabatic conditions and accelerate, the DSBS introduces a modification to the original Hamiltonian $H_0(t)$ as $H_m(t)=H_0(t)+H_c(t)$, so that when driven by $H_m(t)$, the system will evolve along a new path to attain the same target as the desired adiabatic path does.

Assume $\{|\varphi_k(t)\rangle\}$ are the instantaneous eigenstates of $H_0(t)$ with eigenvalues $\{E_k(t)\}$:
$H_0(t)|\varphi_k(t)\rangle=E_k(t)|\varphi_k(t)\rangle$.
First, we define a picture transformation $U(t)=\sum_k|\varphi_k\rangle\langle\varphi_k(t)|$, where $\{|\varphi_k\rangle\}$ is a set of time-independent states independent of $\{|\varphi_k(t)\rangle\}$. $U(t)$ is designed to move to a frame where the adiabatic states $\{|\varphi_k(t)\rangle\}$ are time-independent.
In the adiabatic picture defined by $U(t)$, the modified Hamiltonian becomes
 \begin{align}\label{eqb1}
     H_U(t)&=U(t)H_0(t)U^{\dagger}(t)+U(t)H_c(t)U^{\dagger}(t)+i\dot{U}(t)U^{\dagger}(t)\cr
  &=\sum_kE_k(t)|\varphi_k\rangle\langle\varphi_k|+U(t)H_c(t)U^{\dagger}(t)+i\dot{U}(t)U^{\dagger}(t).
  \end{align}
Second, we define another unitary transformation $V(t)=\sum_k|\tilde{\xi}_k\rangle\langle\tilde{\xi}_k(t)|$,
where $\{|\tilde{\xi}_k(t)\rangle\}$ is a set of time-dependent dressed states in the adiabatic picture, $\{|\tilde{\xi}_k\rangle\}$ is a set of time-independent states. Then, moving from the adiabatic picture to the new  picture defined by $V(t)$, we have
\begin{align}\label{eqb2}
     H_V(t)&=V(t)H_U(t)V^{\dagger}(t)+i\dot{V}(t)V^{\dagger}(t)\cr
  &=V(t)H_{0}(t)V^{\dagger}(t)+V(t)U(t)H_c(t)U^{\dagger}(t)V^{\dagger}(t)\cr
  &+iV(t)\dot{U}(t)U^{\dagger}(t)V^{\dagger}(t)+i\dot{V}(t)V^{\dagger}(t).
  \end{align}
According to DSBS, if an appropriate pair of operators $\{H_c(t),V(t)\}$ is selected to make $H_V(t)$ be diagonal, i.e., $H_V(t)=\sum_k\alpha_k(t)|\tilde{\xi}_k\rangle\langle\tilde{\xi}_k|$, it means that in the adiabatic picture the transitions among $\{|\tilde{\xi}_k(t)\rangle\}$ are inhibited. Then, back to the original picture, the transitions among states $\{|\xi_k(t)\rangle=U^{\dagger}(t)|\tilde{\xi}_k(t)\rangle\}$ are inhibited too.
Thus, if we choose $|\xi_k(t)\rangle$ as the evolution path to realize the desired goal, a shortcut is built up. To ensure the desired goal, the unitary operator $V(t)$ and the evolution state $|\xi_k(t)\rangle$ should satisfy: $V(t_i)=V(t_f)=1$, $|\xi_k(t_i)\rangle=|\psi_i\rangle$, $|\xi_k(t_f)\rangle=|\psi_f\rangle$, where $t_i$ and $t_f$ are the initial and final time, respectively.
 Note that, when $V(t)=1$ and $H_c(t)=i\dot{U}(t)U^{\dagger}(t)$, according to Eq.~(1), $H_U(t)$ is diagonal in the adiabatic picture. In this case,  the shortcut evolution state is $|\varphi_k(t)\rangle$. And this is just the shortcut constructed by transitionless quantum driving \cite{BerryJPA0942,Demirplak0308,ChenPRL10105}. Ref.~\cite{MugaAAMOP1362} has pointed out, this kind of $H_c(t)$ sometimes might cause problems in realizing the completed Hamiltonian $H_m(t)$ in practice.

 Consider the Hamiltonian of a three-level quantum system is $(\hbar=1)$
  \begin{align}\label{eqb3}
     H_0(t)=\Omega_1(t)|1\rangle\langle3|+\Omega_2(t)|2\rangle\langle3|+\textmd{H.c.},
  \end{align}
where $\Omega_1(t)$ and $\Omega_2(t)$ are arbitrary real functions of time. Parameterizing $\Omega_{1}(t)$ and $\Omega_{2}(t)$ as $\Omega_{1}(t)=\Omega(t)\sin\theta(t)$ and $\Omega_{2}(t)=\Omega(t)\cos\theta(t)$ with $\Omega(t)=\sqrt{\Omega_1^2+\Omega_2^2}$, the instantaneous eigenstates of $H_0(t)$ are
\begin{align}\label{eqb4}
    |\varphi_0(t)\rangle&=\cos\theta(t)|1\rangle-\sin\theta(t)|2\rangle,\cr
    |\varphi_{\pm}(t)\rangle&=\frac{1}{\sqrt{2}}[\sin\theta(t)|1\rangle+\cos\theta(t)|2\rangle\pm|3\rangle].
  \end{align}
We define two picture transformations: $U(t)=\sum_{n=0,\pm}|\varphi_n\rangle\langle\varphi_n(t)|$ and $V(t)=\exp[i\beta(t)M_x]$, where $\{|\varphi_n\rangle\}$ is a set of time-independent states, $M_x=(|\varphi_{-}\rangle-|\varphi_{+}\rangle)\langle\varphi_0|/\sqrt{2}$, and $\beta(t)$ is a time-dependent parameter.
Then following the shortcut construction of DSBS, we could inversely design a shortcut as
\begin{align}\label{eqb5}
H_m(t)&=H_0(t)+H_c(t)\cr
&=\Omega_{1m}(t)|1\rangle\langle3|+\Omega_{2m}(t)|2\rangle\langle3|+\textmd{H.c.},
 \end{align}
where
  \begin{align}\label{eqb6}
 \Omega_{1m}(t)&=-\dot{\theta}\sin\theta\cot\beta-\dot{\beta}\cos\theta, \cr
 \Omega_{2m}(t)&=-\dot{\theta}\cos\theta\cot\beta+\dot{\beta}\sin\theta. \end{align}
The corresponding evolution states of the system are
\begin{align}\label{eqb7}
|\xi_0(t)\rangle&=\cos\theta\cos\beta|1\rangle+i\sin\beta|3\rangle-\sin\theta\cos\beta|2\rangle,\cr
|\xi_1(t)\rangle&=\frac{1}{\sqrt{2}}e^{i\varsigma t}[(\sin\theta+i\cos\theta\sin\beta)|1\rangle+i\sqrt{2}\cos\beta|3\rangle \cr
&+(\cos\theta-i\sin\theta\sin\beta)|2\rangle],\cr
|\xi_2(t)\rangle&=\frac{1}{\sqrt{2}}e^{-i\varsigma t}[(\sin\theta-i\cos\theta\sin\beta)|1\rangle-i\sqrt{2}\cos\beta|3\rangle \cr
&+(\cos\theta+i\sin\theta\sin\beta)|2\rangle],
\end{align}
where $\varsigma=-\int_{0}^{t}\frac{\dot{\theta}}{\sin\beta}dt^{\prime}$.
The derivation of Eqs. (5-7) is listed in Appendix A.

In short, for a three-level system whose Hamiltonian possesses the similar form as that in Eq.~(3), we could inversely design the shortcut by Eq.~(5), and the corresponding evolution state could be either one of $\{|\xi_k(t)\rangle\}$ in Eq.~(7).

\section{INVERSE DESIGN SHORTCUT TO IMPLEMENT A Three-qubit nonadiabatic Holonomic GATE}\label{section:III}
\subsection{Theoretical model and the effective Hamiltonian}\label{A}

\begin{figure}[!htb]
 \scalebox{0.16}{\includegraphics{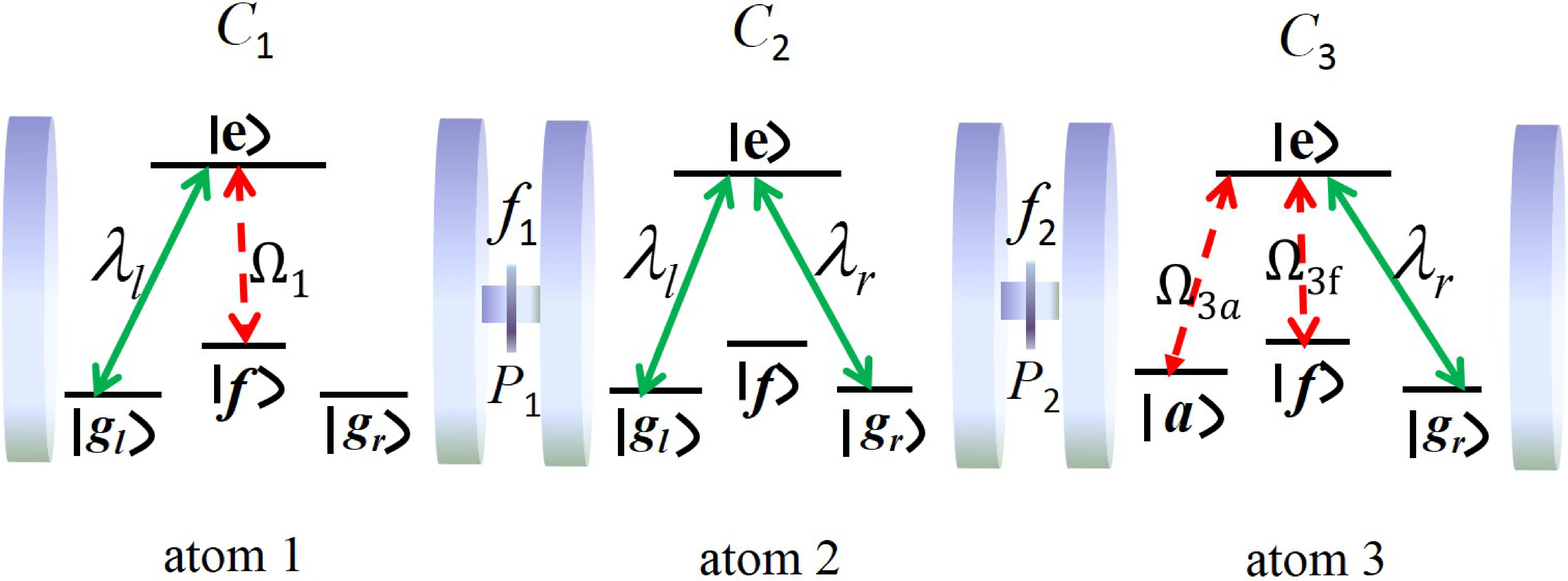}}
   \caption{
         Schematic setup of cavity-atom-fiber combined system. Three identical atoms $1$, $2$, and $3$ are respectively trapped
in three separated cavities $C_1$, $C_2$, and $C_3$, which are linked by two fibers $f_1$ and $f_2$, respectively.
          }
 \label{fig1}
\end{figure}

 The schematic setup for constructing a three-qubit nonadiabatic holonomic gate is shown in
Fig.~1. Three identical atoms (denoted as $1$, $2$, and $3$) are trapped in three distributed linearly arranged optical cavities $C_1$, $C_2$, and $C_3$, respectively. The cavities are bi-mode and connected by two short optical fibers $f_{1}$ and $f_{2}$, in which polarizers $P_{1}$ and $P_{2}$ are inserted in close proximity to $C_2$, respectively. The $P_1$ allows the left-circularly photon to pass and forbids the right-circularly photon, while the $P_2$ allows the right-circularly photon to pass and forbids the left-circularly photon \cite{SongEPL0780,ShanJMO1562}.
Atom 1 and atom 2 have one excited state $|e\rangle$ and three ground states $|g_l\rangle$, $|g_r\rangle$, and $|f\rangle$. Atom 3 has one excited state $|e\rangle$ and three ground states $|g_r\rangle$, $|a\rangle$, and $|f\rangle$. The atomic transition $|g_{l}\rangle_{1(2)}\leftrightarrow|e\rangle_{1(2)}~(|g_{r}\rangle_{3(2)}\leftrightarrow|e\rangle_{3(2)})$
is coupled resonantly to the left-circularly~(right-circularly)
polarized cavity mode with coupling $\lambda_{l}~(\lambda_r)$.
Atomic transition
$|f\rangle_{1}\leftrightarrow|e\rangle_{1}$ is resonantly driven by classical field
$\Omega_{1}(t)$, and atomic transition
$|a\rangle_{3}~(|f\rangle_{3})\leftrightarrow|e\rangle_{3}$ is resonantly driven by $\Omega_{3a}(t)~(\Omega_{3f}(t))$.

In the short-fiber limit, i.e., $L\nu/(2\pi c)\ll1$ ($L$ is the length of the fibers, $\nu$ is the decay rate of the cavity field into a continuum of fiber modes and $c$ is the speed of light), only one resonant mode of the fiber interacts with the cavity mode \cite{SerafiniPRL0696}. Then,
under the rotating-wave
approximation~(RWA), the interaction Hamiltonian for this system
reads ($\hbar=1$)
\begin{align}\label{eqc1}
 & H_{I}=H_{I0}+H_{al},~H_{I0}=H_{ac}+H_{cf}, \cr
 &   H_{al}=\Omega_{1}(t)|e\rangle_{1}\langle f|+\Omega_{3f}(t)|e\rangle_{3}\langle f|+\Omega_{3a}(t)|e\rangle_{3}\langle a|+\textmd{H.c.} ,\cr
 &   H_{ac}=\sum_{k=1}^{2}\lambda_{lk}|e\rangle_{k}\langle g_{l}|a_{lk}
    +\sum_{k=2}^{3}\lambda_{rk}|e\rangle_{k}\langle g_{r}|a_{rk}+\textmd{H.c.},\cr
 &H_{cf}=\upsilon_1b_1^{\dag}(a_{l1}+a_{l2})+\upsilon_2b_2^{\dag}(a_{r1}+a_{r2})+\textmd{H.c.},
    \end{align}
where $a_{lk}~(a_{rk})$ is the annihilation operator for the
left-circularly~(right-circularly) polarized mode of cavity $C_k$, and $b_{1(2)}^{\dag}$ is the creation operator of the resonant mode of fiber $f_{1(2)}$.
For simplicity, we assume $\lambda_{l1}=\lambda_{l2}=\lambda_{r1}=\lambda_{r2}=\lambda$ and $\upsilon_1=\upsilon_2=\upsilon$.

In the scheme, the quantum information is encoded in the subspace $\{|g_l\rangle_1,|g_r\rangle_1,|g_l\rangle_2,|g_r\rangle_2,|a\rangle_3,|f\rangle_3\}$ of the atoms 1, 2, and 3.
For brevity, hereafter, we use the simplified notations $|x,y,z\rangle$ for $|x,y,z\rangle_{1,2,3}\otimes|0\rangle_{\textmd{cav}}\otimes|0\rangle_{\textmd{fib}}$, and $|x,y,z\rangle|1_h\rangle_{s}$ for only cavity $s$ or  fiber $s$ containing an $h$-circularly polarized photon.
Therefore, the computational basis states are spanned by $\mathcal{Z}_0=\{|g_{l},g_{l},a\rangle$, $|g_{l},g_{l},f\rangle,$
$|g_{l},g_{r},a\rangle,$
$|g_{l},g_{r},f\rangle,$
$|g_{r},g_{l},a\rangle,$
$|g_{r},g_{l},f\rangle,$
$|g_{r},g_{r},a\rangle,$
$|g_{r},g_{r},f\rangle\}$.
We assume $\Omega_{3f}(t)=\Omega_3(t)\sin\vartheta$ and $\Omega_{3a}(t)=\Omega_3(t)\cos\vartheta$, where $\vartheta$ is supposed to be a time-independent parameter. Defining $|+\rangle_3=\sin\vartheta|f\rangle_3+\cos\vartheta|a\rangle_3$ and
$|-\rangle_3=\cos\vartheta|f\rangle_3-\sin\vartheta|a\rangle_3$,
Hamiltonian $H_{al}$ is simplified as
\begin{align}\label{eqc2}
H_{al}=\Omega_{1}(t)|e\rangle_{1}\langle f|+\Omega_{3}(t)|e\rangle_{3}\langle +|+\textmd{H.c.}.
\end{align}
Therefore, it is convenient to investigate the gate scheme in the computational basis states spanned by $\mathcal{Z}_{\pm}=\{\mathcal{Z}_{+},\mathcal{Z}_{-}\}$ with $\mathcal{Z}_{+}(\mathcal{Z}_{-})=\{|x,y,+(-)\rangle,~(x,y=g_l,g_r)\}$.
Obviously, the basis states in subspace $\mathcal{Z}_{-}$
are decoupled from Hamiltonian $H_I$, they do not evolve.
 If the system is initially in subspace $\mathcal{Z}_{+}$, when the condition $\Omega_{1(3)}(t)\ll (\lambda,\upsilon)$ is fulfilled, the effective Hamiltonian is derived as
\begin{align}
    H_{\textmd{eff}}=\tilde{\Omega}_{3}(t)|g_{l},g_{r},+\rangle\langle \psi_{3}|+\tilde{\Omega}_{1}(t)|f,g_{l},g_{r}\rangle\langle\psi_{3}|+\textmd{H.c.},
    \end{align}
with the effective Rabi frequency
\begin{align}
    \tilde{\Omega}_{1(3)}(t)=N_3\Omega_{1(3)}(t), ~~N_3=\frac{\upsilon}{\sqrt{3\upsilon^2+2\lambda^2}},
    \end{align}
and
\begin{align}
|\psi_3\rangle&=N_3(|g_{l},g_{r},e\rangle-\frac{\lambda}{\nu}|g_{l},g_{r},g_{r}\rangle|1_r\rangle_{f_2}+|g_{l},e,g_{r}\rangle \cr
&-\frac{\lambda}{\nu}|g_{l},g_{l},g_{r}\rangle|1_l\rangle_{f_1}+|e,g_{l},g_{r}\rangle).
\end{align}
The derivation of Eqs.~(10-12) is listed in Appendix B.

  Since $H_{\textmd{eff}}$ in Eq.~(10) has the similar form as the Hamiltonian in Eq.~(3), we could use the shortcut method raised in section II to inversely design the pulses $\Omega_{1(3)}(t)$ as
\begin{align}
  \Omega_{1}(t)&=\frac{1}{N_3}(\dot{\theta}\cos\theta\cot\beta-\dot{\beta}\sin\theta), \cr
  \Omega_{3}(t)&=\frac{1}{N_3}(\dot{\theta}\sin\theta\cot\beta+\dot{\beta}\cos\theta).
 \end{align}

  \subsection{Implementation of a one-step three-qubit nonadiabatic holonomic gate}\label{B}

  In the following, we assume the initial time is $t=0$ and the final time is $t=T$.
     If we could realize a quantum gate operation
     \begin{eqnarray}
    U_g=\left[
    \begin{array}{ccccccccccc}
    1  &  0 &  0&  0&  0&  0&  0&  0 \\
    0  &  1 &  0&  0&  0&  0&  0&  0 \\
    0  &  0 &  1&  0&  0&  0&  0&  0 \\
    0  &  0 &  0&  1&  0&  0&  0&  0 \\
    0  &  0 &  0&  0&  1&  0&  0&  0 \\
    0  &  0 &  0&  0&  0&  1&  0&  0 \\
    0  &  0 &  0&  0&  0&  0&  1&  0 \\
    0  &  0 &  0&  0&  0&  0&  0&  -1
    \end{array}
    \right],
     \end{eqnarray}
     in subspace $\mathcal{Z}_{\pm}$ with the basis order
     $\{|g_{l},g_{l},-\rangle$, $|g_{l},g_{l},+\rangle,$
$|g_{r},g_{l},-\rangle,$
$|g_{r},g_{l},+\rangle,$
$|g_{r},g_{r},-\rangle,$
$|g_{r},g_{r},+\rangle,$
$|g_{l},g_{r},-\rangle,$
$|g_{l},g_{r},+\rangle\}$
at $t=T$, then back to the original subspace $\mathcal{Z}_{0}$ with basis order $\{|g_{l},g_{l},a\rangle$, $|g_{l},g_{l},f\rangle,$
$|g_{r},g_{l},a\rangle,$
$|g_{r},g_{l},f\rangle,$
$|g_{r},g_{r},a\rangle,$
$|g_{r},g_{r},f\rangle,$
$|g_{l},g_{r},a\rangle,$
$|g_{l},g_{r},f\rangle\}$, we obtain
 \begin{eqnarray}
    U_g(\vartheta)=\left[
    \begin{array}{ccccccccccc}
    1  &  0 &  0&  0&  0&  0&  0&  0 \\
    0  &  1 &  0&  0&  0&  0&  0&  0 \\
    0  &  0 &  1&  0&  0&  0&  0&  0 \\
    0  &  0 &  0&  1&  0&  0&  0&  0 \\
    0  &  0 &  0&  0&  1&  0&  0&  0 \\
    0  &  0 &  0&  0&  0&  1&  0&  0 \\
    0  &  0 &  0&  0&  0&  0&  -\cos 2\vartheta  &  -\sin 2\vartheta \\
    0  &  0 &  0&  0&  0&  0&  -\sin 2\vartheta  &  \cos 2\vartheta
    \end{array}
    \right].
     \end{eqnarray}
 For $\vartheta=-\frac{\pi}{4}$, $U_g(\vartheta)$ becomes
   \begin{eqnarray}
    U_g(-\frac{\pi}{4})=\left[
    \begin{array}{ccccccccccc}
    1  &  0 &  0&  0&  0&  0&  0&  0 \\
    0  &  1 &  0&  0&  0&  0&  0&  0 \\
    0  &  0 &  1&  0&  0&  0&  0&  0 \\
    0  &  0 &  0&  1&  0&  0&  0&  0 \\
    0  &  0 &  0&  0&  1&  0&  0&  0 \\
    0  &  0 &  0&  0&  0&  1&  0&  0 \\
    0  &  0 &  0&  0&  0&  0&  0&  1 \\
    0  &  0 &  0&  0&  0&  0&  1&  0
    \end{array}
    \right],
     \end{eqnarray}
     which is a Toffoli gate \cite{Toffoli,ShaoPLA09374}.
     When $\vartheta=\frac{\pi}{2}$, a controlled three-qubit phase gate is realized
      \begin{eqnarray}
    U_g(\frac{\pi}{2})=\left[
    \begin{array}{ccccccccccc}
    1  &  0 &  0&  0&  0&  0&  0&  0 \\
    0  &  1 &  0&  0&  0&  0&  0&  0 \\
    0  &  0 &  1&  0&  0&  0&  0&  0 \\
    0  &  0 &  0&  1&  0&  0&  0&  0 \\
    0  &  0 &  0&  0&  1&  0&  0&  0 \\
    0  &  0 &  0&  0&  0&  1&  0&  0 \\
    0  &  0 &  0&  0&  0&  0&  1&  0 \\
    0  &  0 &  0&  0&  0&  0&  0&  -1
    \end{array}
    \right].
     \end{eqnarray}

  To obtain the gate operation $U_g$ shown in Eq.~(14), we analyse the dynamics of the system. Since states in $\mathcal{Z}_{-}$ are decoupled from $H_{I}$, they do not evolve.
    For the states in subspace $\mathcal{Z}_{+}$, when $\Omega_{1(3)}(t)\ll (\lambda,\upsilon)$, according to the effective Hamiltonian $H_{\textmd{eff}}$ in Eq.~(10), only the state $|g_l,g_r,+\rangle$ will evolve.
         As $H_{\textmd{eff}}$ possesses the similar form as the Hamiltonian in Eq.~(3), according to section II, we can choose $|\xi_0(t)\rangle$ as the evolution state of the system. Here, $|\xi_0(t)\rangle$ has the form as
         \begin{align}
      |\xi_0(t)\rangle&=\cos\theta\cos\beta|g_{l},g_{r},+\rangle-\sin\theta\cos\beta|f,g_{l},g_{r}\rangle \cr
      &+i\sin\beta|\psi_{3}\rangle.
 \end{align}
By setting the boundary conditions:
\begin{align}
 \theta(0)=0,\theta(T)=-\pi,\beta(0)=0,\beta(T)=0,
 \end{align}
 we obtain $|\xi_0(0)\rangle=|g_{l},g_{r},+\rangle$ and $|\xi_0(T)\rangle=-|g_{l},g_{r},+\rangle$ in time interval $[0,T]$.
   That is, by designing pulses in Eq.~(13) with the boundary conditions in Eq.~(19), the quantum gate $U_g$ given in Eq.~(18) is realizable.

Besides, according to the above discussion, it is not difficult to find that the conditions (i-ii) are satisfied in time interval $[0,T]$:
 \begin{align}
 &(i)~\sum\limits_{l=1}^{8}|\chi_l(T)\rangle\langle\chi_l(T)|=\sum\limits_{l=1}^{8}|\chi_l(0)\rangle\langle\chi_l(0)|,
\cr
&(ii)~\langle\chi_l(t)|H_{\textmd{eff}}(t)|\chi_{l^{\prime}}(t)\rangle=0, ~(l,l^{\prime}=1,...,8),
 \end{align}
where $|\chi_1(t)\rangle=|\xi_0(t)\rangle$ and $\{|\chi_l(t)\rangle~(l=2,...,7)\}$ are the states in subspace $\mathcal{Z}_{\pm}$ except the state $|g_{l},g_{r},+\rangle$.

 In the following, as an illustration, we design pulses with the boundary conditions shown in Eq.~(19). For simplicity, we choose $\lambda=\upsilon$ for discussion. According to Eq.~(13), the inversely designed pulses are
\begin{align}
 \Omega_{1}(t)&=\sqrt{5}(\dot{\theta}\cos\theta\cot\beta-\dot{\beta}\sin\theta), \cr
 \Omega_{3}(t)&=\sqrt{5}(\dot{\theta}\sin\theta\cot\beta+\dot{\beta}\cos\theta). \end{align}
To satisfy the
boundary conditions shown in Eq. (19) as well as avoid the singularity of the expression for each pulse, as an example, we select Trigonometric functions to fit the boundary conditions and the
corresponding $\theta(t)$ and $\beta(t)$ are set as
\begin{align}
 \theta(t)&=-\pi[2+\cos(\frac{\pi t}{T})]\sin^4(\frac{\pi t}{2T}), \cr
 \beta(t)&=A\sin^2(\frac{2\pi t}{T}),
  \end{align}
where $A$ is a time-independent parameter which controls the maximal value of $\beta(t)$. If we set $0<A<\pi/2$, when $A$ decreases, according to Eqs.~(18) and (22), the population of the intermediate state $|\psi_3\rangle$ also decreases, which may be useful for depressing the effects of dissipation. On the other hand,
according to  Eqs.~(21) and (22), we know that $\Omega_{1(3)}(t)$ is strongly related with $T$ and $A$. To well satisfy the condition $(\lambda,\upsilon)\gg\Omega_{1(3)}(t)$ and choose apposite $T$ and $A$, we investigate the fidelity with $T$ and $A$, respectively.
For a quantum gate operation, it is important that the gate fidelity is independent of the input state in the strict sense, therefore, we use the average effective fidelity \cite{YinPRA0775,SuPRA1693}
\begin{eqnarray}
 \overline{F}_e=\frac{1}{8\pi^3}\int_0^{2\pi}d\mu_1\int_0^{2\pi}d\mu_2\int_0^{2\pi}d\mu_3F_e(\mu_1,\mu_2,\mu_3),
  \end{eqnarray}
with the fidelity for the effective dynamics defined as
\begin{align}
 F_e(\mu_1,\mu_2,\mu_3)\!=|\langle\Psi_T(\mu_1,\mu_2,\mu_3)|U_e(T)|\Psi_0(\mu_1,\mu_2,\mu_3)\rangle|^2,\cr
  \end{align}
where
\begin{align}
&|\Psi_0(\mu_1,\mu_2,\mu_3)\rangle \cr
&=\sin\mu_1\sin\mu_2(\sin\mu_3|g_l,g_r,+\rangle+\cos\mu_3|g_l,g_r,-\rangle)\cr
&+\sin\mu_1\cos\mu_2(\sin\mu_3|g_r,g_r,+\rangle+\cos\mu_3|g_r,g_r,-\rangle)\cr
&+\cos\mu_1\sin\mu_2(\sin\mu_3|g_l,g_l,+\rangle+\cos\mu_3|g_l,g_l,-\rangle)\cr
&+\cos\mu_1\cos\mu_2(\sin\mu_3|g_l,g_r,+\rangle+\cos\mu_3|g_l,g_r,-\rangle),\cr
\end{align}
is the initial state,
\begin{align}
&|\Psi_T(\mu_1,\mu_2,\mu_3)\rangle \cr
&=\sin\mu_1\sin\mu_2(-\sin\mu_3|g_l,g_r,+\rangle+\cos\mu_3|g_l,g_r,-\rangle)\cr
&+\sin\mu_1\cos\mu_2(\sin\mu_3|g_r,g_r,+\rangle+\cos\mu_3|g_r,g_r,-\rangle)\cr
&+\cos\mu_1\sin\mu_2(\sin\mu_3|g_l,g_l,+\rangle+\cos\mu_3|g_l,g_l,-\rangle)\cr
&+\cos\mu_1\cos\mu_2(\sin\mu_3|g_l,g_r,+\rangle+\cos\mu_3|g_l,g_r,-\rangle),\cr
\end{align}
is the final state after the gate operation $U_g$, and
$U_e(T)$ is the evolution operator of the effective dynamics with the designed Rabi frequencies $\Omega_{1(3)}$ in Eq.~(21).

In Fig.~2(a), we plot a contour image for the average effective fidelity $\overline{F}_e$ with the full Hamiltonian $H_I$ in Eq.~(8). As seen from Fig.~2(a), a wide range of $A$ and $\lambda T$ is available for a relatively high $\overline{F}_e$.
To get a high $\overline{F}_e$ with small value of $\lambda T$ and robustness against dissipation, as seen from Fig.~2(b), we choose $A=0.55$ for illustration. Note that, other values are also available to implement the scheme.

With $A=0.55$, we plot the time dependence of the pulses $\Omega_1(t)$ and $\Omega_3(t)$ by dotted-red lines in Fig.~3, respectively. As seen from Fig.~3, when $A=0.55$, the maximum of the pulses $\Omega_{1(3)}(t)$ is $\Omega^{M}=\max\{\Omega_1(t),\Omega_3(t)\}\approx28/T$. Note that, to well execute the scheme, the condition $(\lambda,\upsilon)\gg\Omega_{1(3)}(t)$ should be ensured to obtain the effective Hamiltonian $H_{\textmd{eff}}$, and the condition can be replaced by $(\lambda,\upsilon)\gg\Omega^{M}$. As shown in Fig.~2(b), $\overline{F}_e\sim0.9999$ when $\lambda\geq 120/T$. That means even if the condition $(\lambda,\upsilon)\gg\Omega^M$ is not well satisfied, one also can get a high fidelity with the present scheme. Here, we choose $\lambda=150/T$ for illustration. In realistic experiments, the fabrication of various high-Q microcavities is well developed \cite{ArmaniNAT03421,SpillanePRL0391} and the strong coupling $\lambda\sim2\pi\times750$MHz between atom and cavity is available in a toroidal microcavity \cite{SpillanePRA0571}, and the coupling strength $\upsilon=\lambda$ is possible because $\upsilon$ can be controlled by adjusting the reflectivity of the cavity mirror connected to the fiber \cite{SerafiniPRL0696}. Therefore, the above choice of parameters is available in experiments.

 For the convenience of the experimental realization, we engineer Gaussian and sin-shaped pulses to fit $\Omega_1(t)$ and $\Omega_3(t)$, and obtain two substituted pulses $\bar{\Omega}_1(t)$ and $\bar{\Omega}_3(t)$ plotted by solid-blue lines in Fig.~3, respectively. The expressions of $\bar{\Omega}_1(t)$ and $\bar{\Omega}_3(t)$ are
\begin{align}
 \bar{\Omega}_1(t)&=\frac{3}{T}\sin(\frac{4\pi t}{T})-\frac{20}{T}\sin(\frac{2\pi t}{T}),
 \cr
 \bar{\Omega}_3(t)&=\frac{24.5}{T}\exp{[-(\frac{t-0.5T}{0.22T})^2]}. \end{align}
As seen from Fig.~3, the curve of $\bar{\Omega}_{1(3)}(t)$ coincides well with that of $\Omega_{1(3)}(t)$.
To see the effectiveness of $\bar{\Omega}_{1(3)}(t)$, we also examine $1-F_e(\mu_1,\mu_2,\mu_3=\pi/20)$ versus $\mu_{1(2)}$ and $1-F_e(\mu_1=\pi/20,\mu_2,\mu_3)$ versus $\mu_{2(3)}$ with $\bar{\Omega}_{1(3)}(t)$ in Fig.~4, respectively. As shown in Fig.~4, the infidelity $1-F_e$ is very small and varies little for different $\mu$, which demonstrates that the substitute $\bar{\Omega}_{1(3)}(t)$ used in the proposed shortcut scheme is effective.
Besides, we also show that, with the pulses $\bar{\Omega}_{1}(t)$ and $\bar{\Omega}_{3}(t)$, the basis states in subspace $\mathcal{Z}_{+}$ hardly evolve except state $|g_{l},g_{r},+\rangle$. In the evolution, the population of state $|g_{l},g_{r},+\rangle$ is unchange at $t=0$ and $t=T$, while the phase changes $\pi$. This point also confirms the effectiveness of the present scheme. Detailed descriptions about this point are presented in Appendix C. Moreover, we can numerically show that the present nonadiabatic gate is faster than adiabatic gate. The detailed comparison is given in Appendix D.

\begin{figure}
 \scalebox{0.3}{\includegraphics{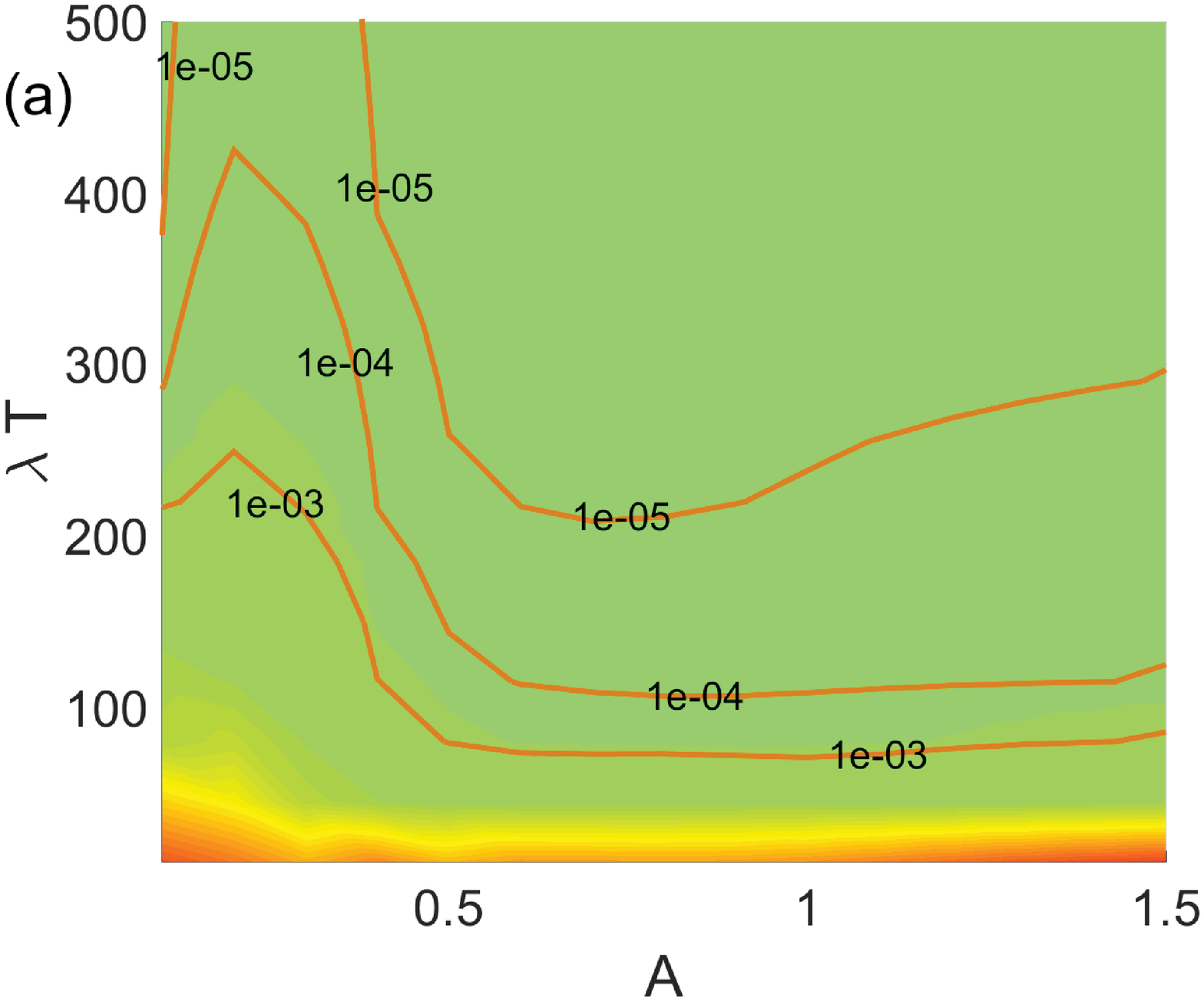}}
 \scalebox{0.3}{\includegraphics{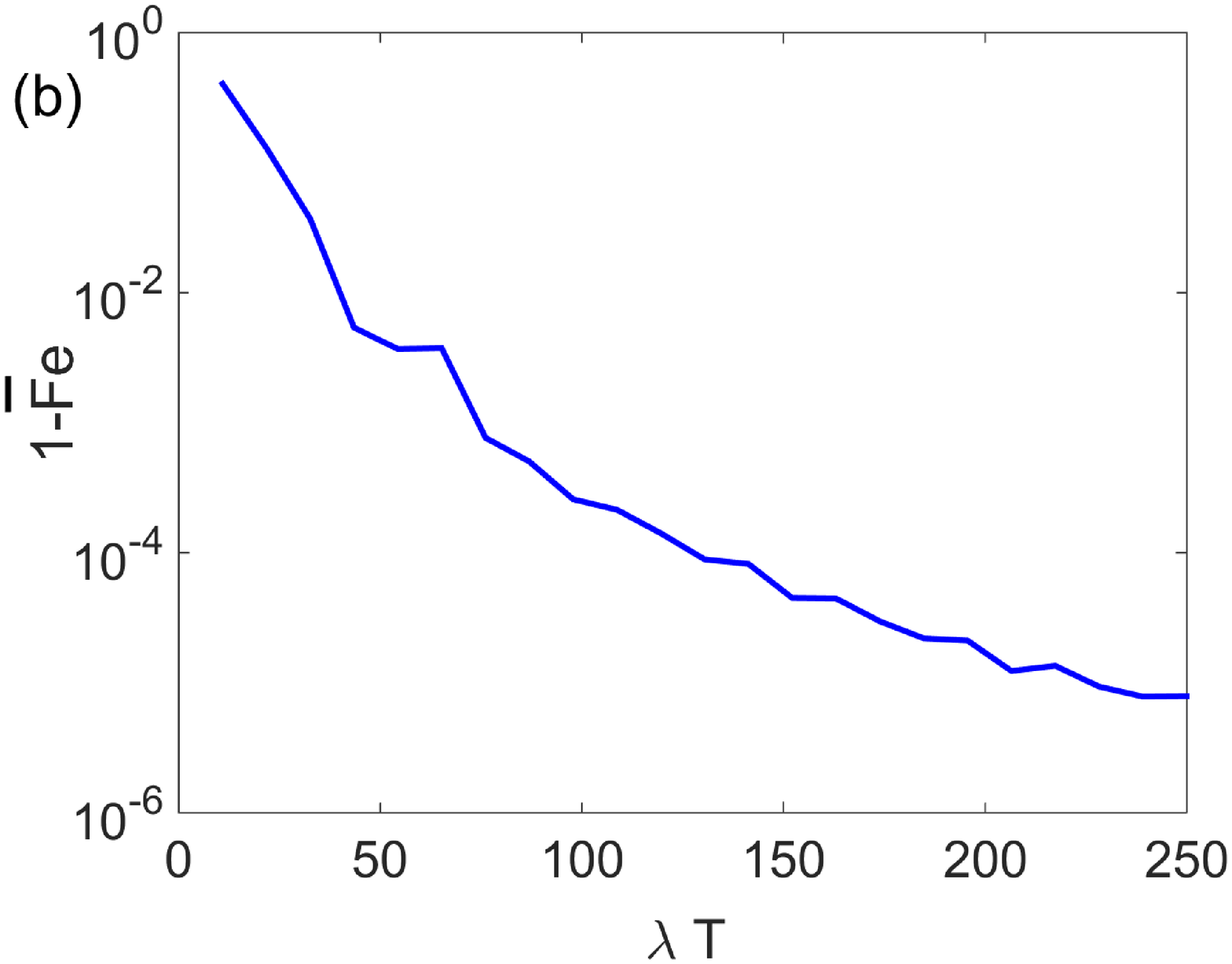}}
  \caption{
         (a) Contour image for the average infidelity $1-\bar{F}_e$ versus $A$ and $\lambda T$.
         (b) The average infidelity $1-\bar{F}_e$ versus $\lambda T$ for $A=0.55$.
          }
 \label{fig2}
\end{figure}

\begin{figure}[!htb]
 \scalebox{0.33}{\includegraphics{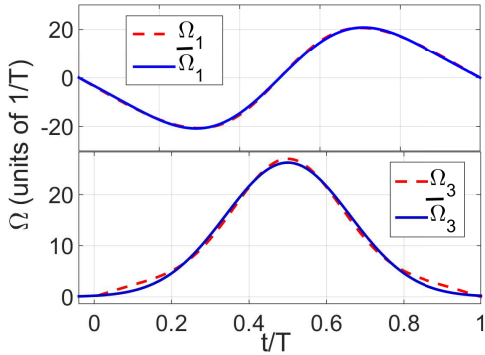}}
    \caption{
          Time dependence of $\Omega_{1(3)}(t)$ and $\bar{\Omega}_{1(3)}(t)$.
            }
 \label{fig3}
\end{figure}

\section{Numerical simulation and discussion}\label{section:IV}

\begin{figure}
 \scalebox{0.3}{\includegraphics{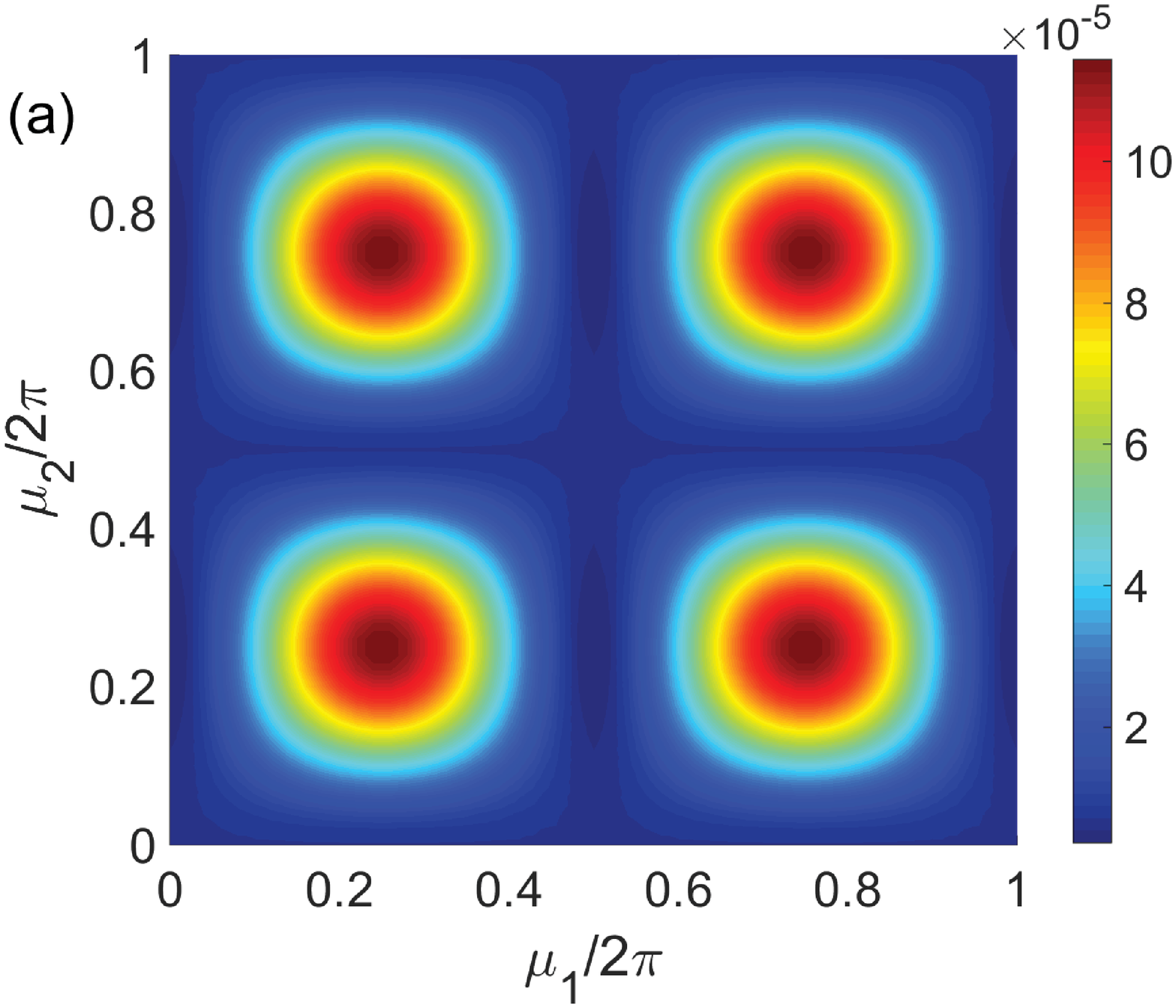}}
 \scalebox{0.3}{\includegraphics{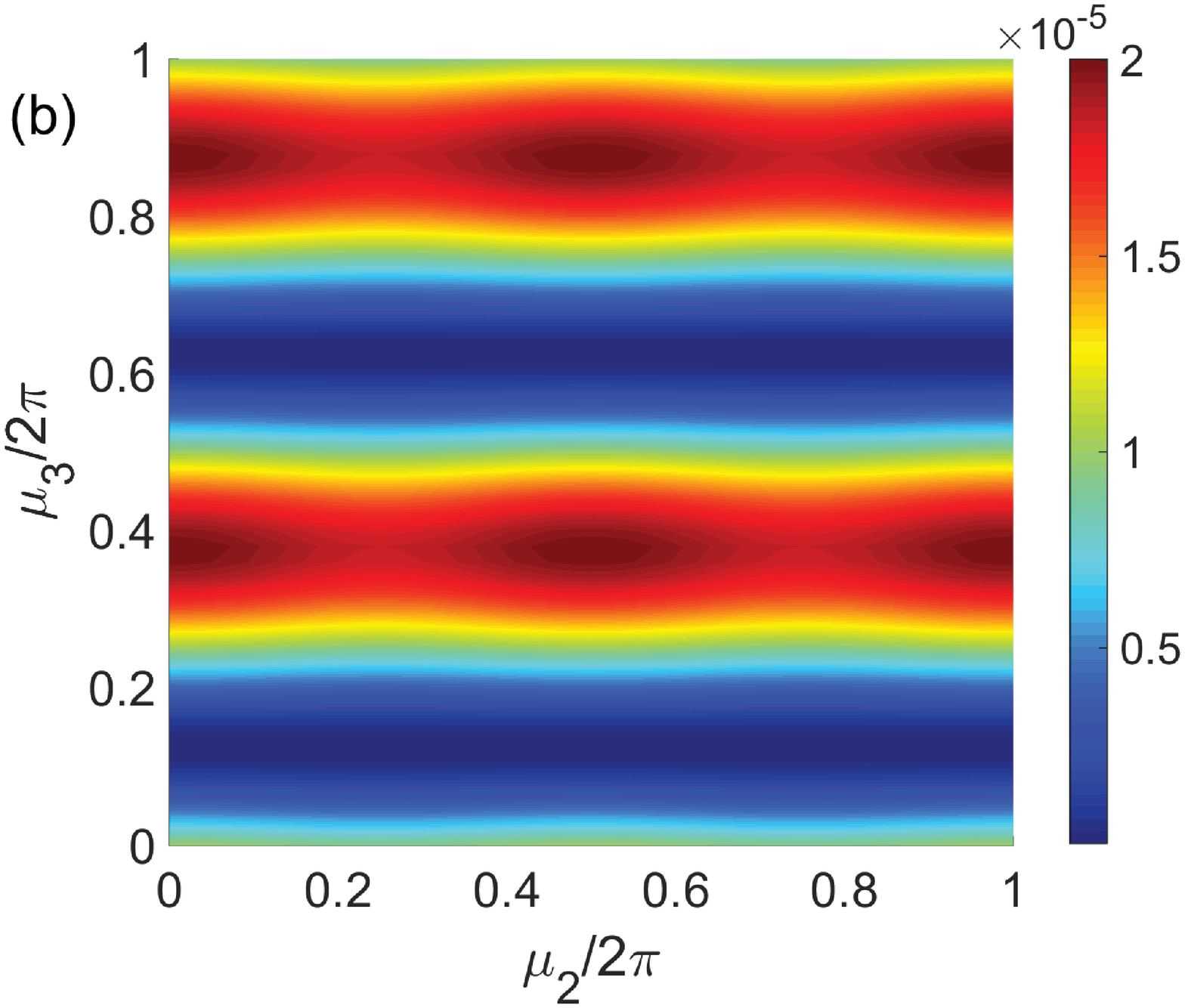}}
   \caption{
          The infidelity $1-F_e(\mu_1,\mu_2,\mu_3)$ versus (a) $\mu_1$, $\mu_2$, and $\mu_3=\pi/20$, (b)  $\mu_1=\pi/20$, $\mu_2$, and $\mu_3$.
          }
 \label{fig4}
\end{figure}

In this section, we numerically investigate the feasibility of the present gate scheme. We take the three-qubit Toffoli gate ($\vartheta=-\pi/4$) as an example.
The quantum information is encoded in the subspace $\{|g_l\rangle_1,~|g_r\rangle_1,~|g_l\rangle_2,~|g_r\rangle_2,~|a\rangle_3,~|f\rangle_3\}$ and the computational basis states are  $\{|g_{l},g_{l},a\rangle$, $|g_{l},g_{l},f\rangle,$
$|g_{r},g_{l},a\rangle,$
$|g_{r},g_{l},f\rangle,$
$|g_{r},g_{r},a\rangle,$
$|g_{r},g_{r},f\rangle,$
$|g_{l},g_{r},a\rangle,$
$|g_{l},g_{r},f\rangle\}$. For the sake of generality, the gate fidelity discussed here is also the average fidelity:
\begin{align}
 &\overline{F}_{to}=\frac{1}{8\pi^3}\int_0^{2\pi}\int_0^{2\pi}\int_0^{2\pi}d\mu_1d\mu_2d\mu_3F_{to}(\mu_1,\mu_2,\mu_3),
  \cr
& F_{to}(\mu_1,\mu_2,\mu_3)=|\langle\Phi_T(\mu_1,\mu_2,\mu_3)|U(T)|\Phi_0(\mu_1,\mu_2,\mu_3)\rangle|^2, \end{align}
where
\begin{align}
&|\Phi_0(\mu_1,\mu_2,\mu_3)\rangle \cr
&=\sin\mu_1\sin\mu_2(\sin\mu_3|g_l,g_r,a\rangle+\cos\mu_3|g_l,g_r,f\rangle)\cr
&+\sin\mu_1\cos\mu_2(\sin\mu_3|g_r,g_r,a\rangle+\cos\mu_3|g_r,g_r,f\rangle)\cr
&+\cos\mu_1\sin\mu_2(\sin\mu_3|g_l,g_l,a\rangle+\cos\mu_3|g_l,g_l,f\rangle)\cr
&+\cos\mu_1\cos\mu_2(\sin\mu_3|g_l,g_r,a\rangle+\cos\mu_3|g_l,g_r,f\rangle),\cr
\end{align}
is the initial state of the system,
\begin{align}
&|\Phi_T(\mu_1,\mu_2,\mu_3)\rangle \cr
&=\sin\mu_1\sin\mu_2(\cos\mu_3|g_l,g_r,a\rangle+\sin\mu_3|g_l,g_r,f\rangle)\cr
&+\sin\mu_1\cos\mu_2(\sin\mu_3|g_r,g_r,a\rangle+\cos\mu_3|g_r,g_r,f\rangle)\cr
&+\cos\mu_1\sin\mu_2(\sin\mu_3|g_l,g_l,a\rangle+\cos\mu_3|g_l,g_l,f\rangle)\cr
&+\cos\mu_1\cos\mu_2(\sin\mu_3|g_l,g_r,a\rangle+\cos\mu_3|g_l,g_r,f\rangle),\cr
\end{align}
denotes the final state after the Toffoli gate operation $U_g(-\frac{\pi}{4})$, and
$U(T)$ is the evolution operator.

In a practical situation, it is natural to explore how robust the scheme is under variations or noise. Thus, in the following,
we will test the scheme with simulations of the realistic situation.
Firstly, we take into account of the fluctuations of the control parameters caused by imperfect operations. In the scheme, the main imperfect variations are respectively $\delta T$, $\delta\bar{\Omega}^{M}_{1(3)}$, $\delta{\nu}$, and $\delta{\lambda}$ acted on $T$, $\bar{\Omega}^{M}_{1(3)}$, $\nu$, and $\lambda$.
Here $\bar{\Omega}^{M}_{1(3)}$ is the abbreviation of $\bar{\Omega}^{M}_{1}$ and $\bar{\Omega}^{M}_{3}$, and $\bar{\Omega}^{M}_{1}$ ($\bar{\Omega}^{M}_{3}$) represents the maximum of the pulse $\bar{\Omega}_{1}$ ($\bar{\Omega}_{3}$). $\delta\bar{\Omega}^{M}_{1}$ ($\delta\bar{\Omega}^{M}_{3}$) denotes the variation to $\bar{\Omega}^{M}_{1}$ ($\bar{\Omega}^{M}_{3}$).
The average infidelities $1-\bar{F}_{to}$ as the functions of $(\delta{\nu}/\nu,~\delta{\lambda}/\lambda)$ and $(\delta T/T,~\delta{\bar{\Omega}^{M}_{1(3)}}/\bar{\Omega}^{M}_{1(3)})$ are shown in Figs.~5(a) and 5(b), respectively. As seen from Fig.~5, it is evident that the scheme is insensitive to typical variations even on the scale of $10\%$ of the parameters $T$, $\nu$, and $\lambda$, respectively. We can see from Fig.~5(b), the amplitudes of the pulses influence a little bit. When the amplitudes of the pulses decrease, the average infidelity $1-\bar{F}_{to}$ increases. Nevertheless, even when the amplitudes change by up to $10\%$, the average infidelity $1-\bar{F}_{to}$ is still $\sim10^{-3}$. Thus, the present Toffoli gate holds a good performance on the disturbance of imperfect experimental operations.
\begin{figure}
 \scalebox{0.3}{\includegraphics{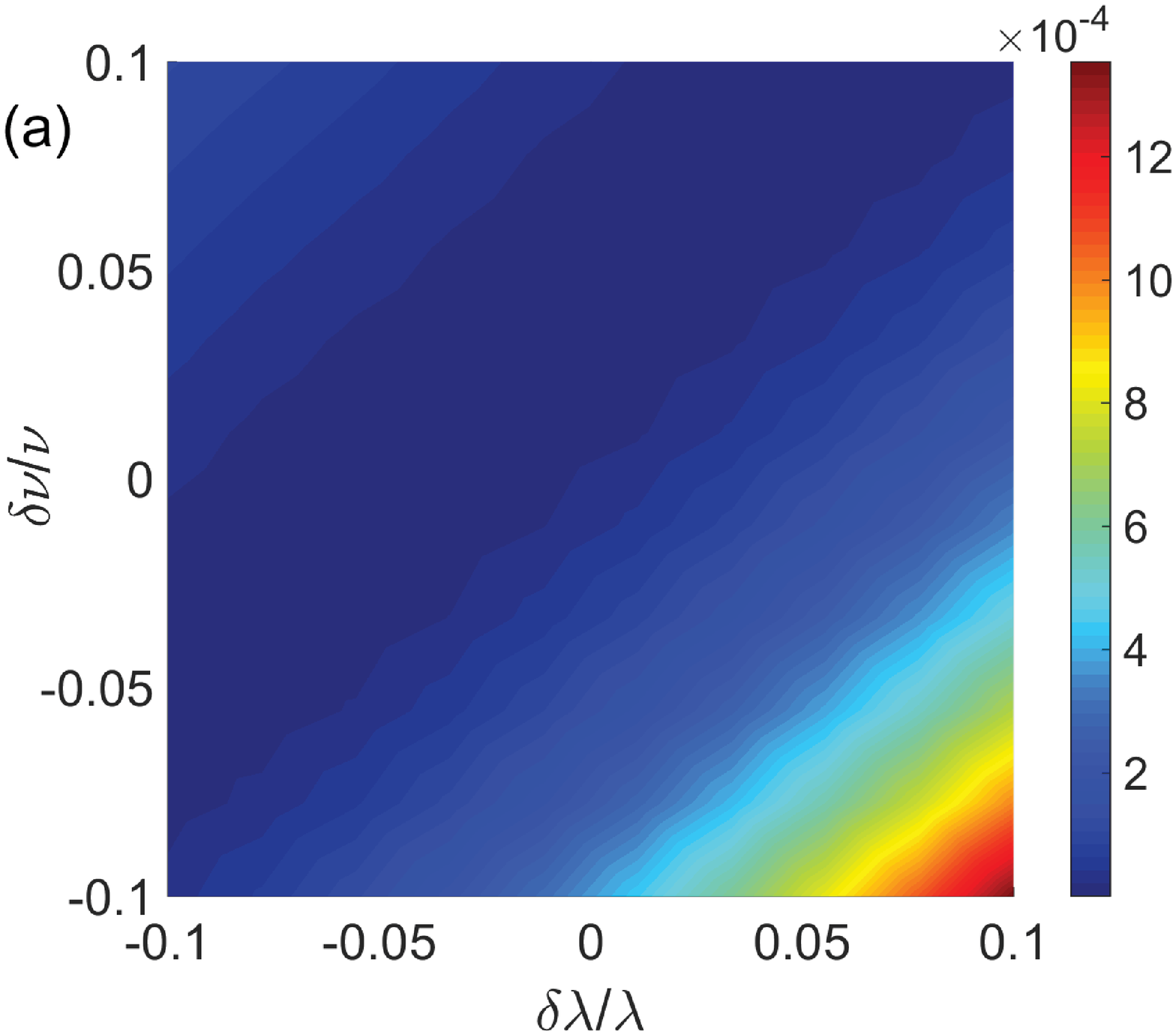}}
 \scalebox{0.3}{\includegraphics{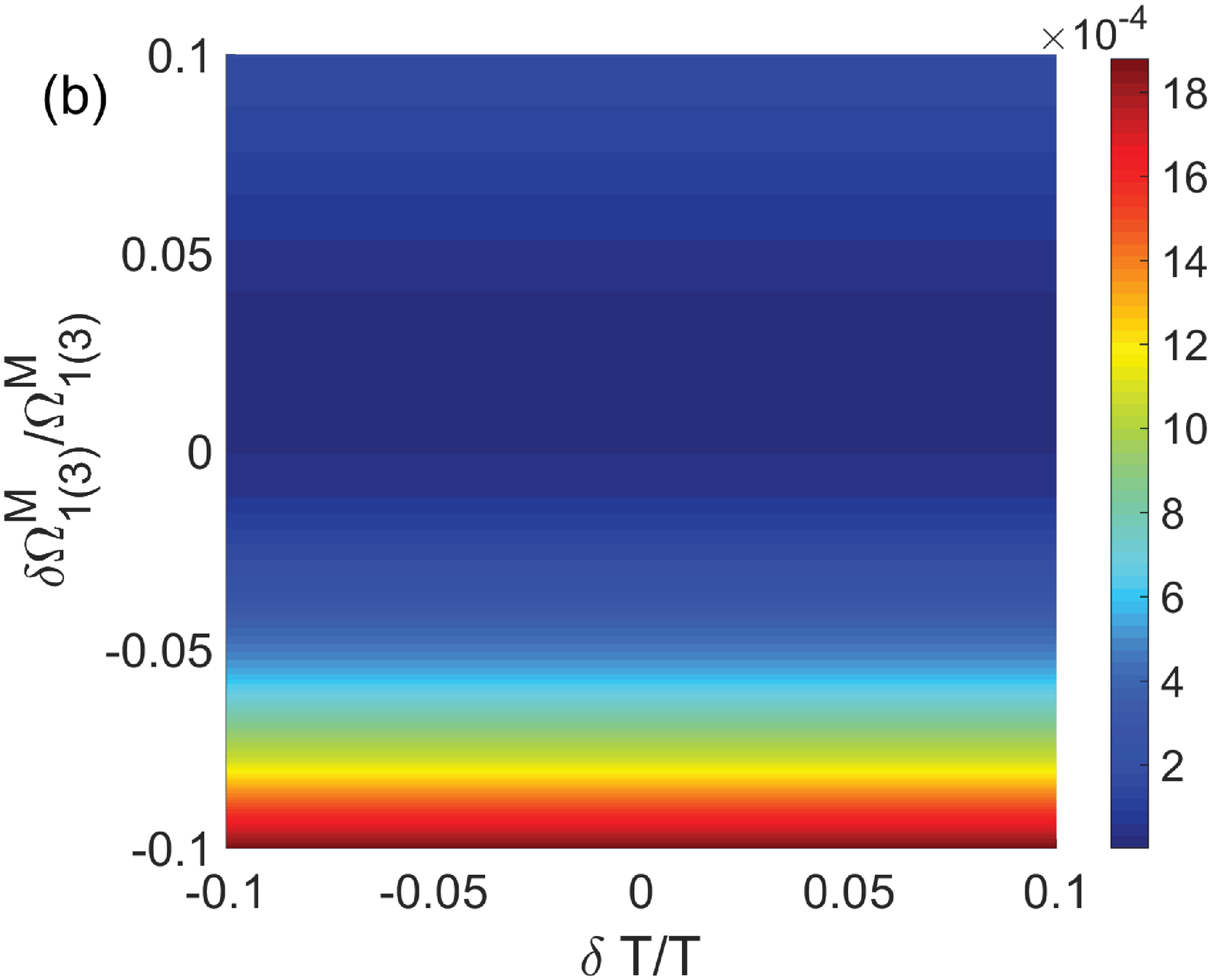}}
  \caption{
          (a) The average infidelity $1-\bar{F}_{to}$ of the Toffoli gate versus $\delta{\nu}/\nu$ and $\delta{\lambda}/\lambda$.
          (b) The average infidelity $1-\bar{F}_{to}$ of the Toffoli gate versus  $\delta{\bar{\Omega}^{M}_{1(3)}}/\bar{\Omega}^{M}_{1(3)}$ and $\delta T/T$.
          }
 \label{fig5}
\end{figure}

Besides imperfections from the control of the quantum system, decoherence, arisen from the inevitable interaction between the quantum system and environment, is another main challenge in implementing robust quantum gates.
 Here, the decoherence channels include the atomic spontaneous emission, cavity decay, and fiber photon leakage. In view of these factors, the whole system is dominated by the master equation in the Lindblad form \cite{Lindblad7648}
 \begin{align}
  \dot{\rho}&=i[\rho, H_{I}(t)]+\sum_{k=1}^{2}\frac{\kappa_{fk}}{2}(2b_{k}\rho b^{\dag}_{k}-b^{\dag}_{k}b_{k}\rho -\rho b^{\dag}_{k}b_{k})\cr
  &+\sum_{i=1}^{3}\sum_{j=f,g_l,g_r,a}\frac{\gamma_{ij}}{2}(2S^{-}_{ij}\rho S^{\dag}_{ij}-S^{\dag}_{ij}S^{-}_{ij}\rho -\rho S^{\dag}_{ij}S^{-}_{ij})\cr
  &+\sum_{k=1}^{2}\frac{\kappa_{ck}}{2}(2a_{lk}\rho a^{\dag}_{lk}-a^{\dag}_{lk}a_{lk}\rho -\rho a^{\dag}_{lk}a_{lk})\cr
&+\sum_{k=2}^{3}\frac{\kappa_{ck}}{2}(2a_{rk}\rho a^{\dag}_{rk}-a^{\dag}_{rk}a_{rk}\rho -\rho a^{\dag}_{rk}a_{rk}),
\end{align}
where $S^{\dag}_{ij}=|e\rangle_i\langle j|$, $S^{-}_{ij}=|j\rangle_i\langle e|$, $\gamma_{ij}$ is the atomic spontaneous emission rate of atoms, and $\kappa_{ck}$ $(\kappa_{fk})$ is the decay rate of the $k$th cavity (fiber). For simplicity, we assume that $\gamma_{ij}=\gamma$, $\kappa_{ck}=\kappa_c$, and $\kappa_{fk}=\kappa_f$.
When the decoherence is considered, the fidelity of the Toffoli gate with initial state $|\Phi_0(\mu_1,\mu_2,\mu_3)\rangle$ is derived as
\begin{align}
  F_{to}(\mu_1,\mu_2,\mu_3)=|\langle\Phi_T(\mu_1,\mu_2,\mu_3)|\rho|\Phi_0(\mu_1,\mu_2,\mu_3)\rangle|^2,
  \end{align}
with $\rho=\rho(\mu_1,\mu_2,\mu_3,T)$ being the density operator of the system at time $T$. The average fidelity $\bar{F}_{to}$ can still be calculated by using Eq.~(28).
In Fig.~6, we display the average fidelity $\bar{F}_{to}$ versus $\gamma/\lambda$, $\kappa_c/\lambda$, and $\kappa_f/\lambda$, respectively. It turns out that: (i) the present
Toffoli gate with this set of parameters is robust. When the dissipation is up to $0.01\lambda$, $\bar{F}_{to}$ is still higher than 0.986. (ii) among the three dissipation factors, the atomic spontaneous emission influences a little bit. This is easy to understand by the effective Hamiltonian $H_{\textmd{eff}}$ in Eq.~(10) and the intermediate state $|\psi_3\rangle$ in Eq.~(12). As seen from Eq.~(12), $|\psi_3\rangle$ contains the excited states of all three atoms, therefore, the
influence from dissipation of the atoms would be greater than the other two factors. To obtain a higher
average fidelity, one possible way is to increase the intensity of the pulses, thus the population of the intermediate state $|\psi_3\rangle$ would decrease.

\begin{figure}[!htb]
 \scalebox{0.3}{\includegraphics{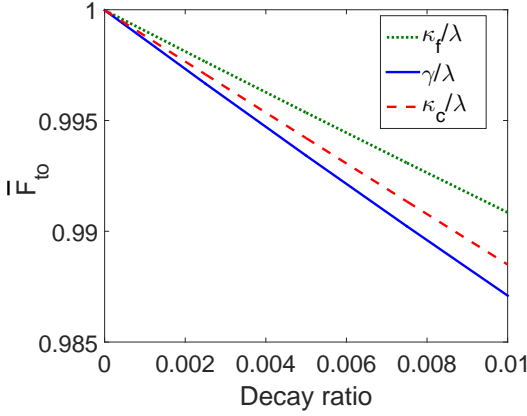}}
   \caption{
         Effects of the decoherence on the average fidelity $\bar{F}_{to}$: each of the three curves denotes the average fidelity versus the corresponding decoherence factor with the other two factors are zero.
          }
 \label{fig6}
\end{figure}

\section{CONCLUSION}\label{section:V}

In this paper, we have raised a scheme for one-step implementation of a three-qubit nonadiabatic holonomic gate by using the shortcuts to adiabaticity (STA).
The scheme combines the virtues of STA and nonadiabatic holonomy.
By simplifying the Hamiltonian of the system to the effective Hamiltonian, two engineering pulses have been designed to fulfill the target. Besides, numerical simulations have been performed, and the results have shown that a high-fidelity nonadiabatic holonomic gate is realizable with a wide range of coupling strength and the interaction time. Further, the numerical results also have demonstrated that the presented quantum gate scheme is robust against the decoherence and operational imperfections. We hope that our
work may open up further investigation towards the implementation of fast and robust quantum computation in the
quantum-information-processing.

\section*{ACKNOWLEDGEMENT}

This work was supported by the National Natural Science Foundation
of China under Grants No. 11575045, No. 11675046, and No. 11747011,
and the Major State Basic Research Development Program of China
under Grant No. 2012CB921601, and the Natural Science Foundation of
Fujian Province under Grant No. 2018J01414.

\appendix

\section{}

In this Appendix A, we show the derivation of Eqs.~(5-7).
According to Eqs.~(3) and (4), the Hamiltonian of the three-level system is
  \begin{align}\label{eqb1}
     H_0(t)=\Omega(t)\sin\theta(t)|1\rangle\langle3|+\Omega(t)\cos\theta(t)|2\rangle\langle3|+\textmd{H.c.},
  \end{align}
and the instantaneous eigenstates of $H_0(t)$ are
\begin{align}\label{eqb2}
    |\varphi_0(t)\rangle&=\cos\theta(t)|1\rangle-\sin\theta(t)|2\rangle,\cr
    |\varphi_{\pm}(t)\rangle&=\frac{1}{\sqrt{2}}[\sin\theta(t)|1\rangle+\cos\theta(t)|2\rangle\pm|3\rangle].
  \end{align}
Transforming to the adiabatic picture defined by $U(t)=\sum_{n=0,\pm}|\varphi_n\rangle\langle\varphi_n(t)|$, where $\{|\varphi_n\rangle\}$ is a set of time-independent states, the Hamiltonian becomes
 \begin{align}\label{eqb3}
    H_{0}^{a}(t)&=U(t)H_0(t)U^{\dag}(t)+i\dot{U}(t)U^{\dag}(t) \cr
    &=\Omega(t)M_z+\dot{\theta}M_y,
  \end{align}
where $M_x=(|\varphi_{-}\rangle-|\varphi_{+}\rangle)\langle\varphi_0|/\sqrt{2}+\textmd{H.c.}$,
$M_y=i(|\varphi_{-}\rangle+|\varphi_{+}\rangle)\langle\varphi_0|/\sqrt{2}+\textmd{H.c.}$,
and $M_z=|\varphi_{+}\rangle\langle\varphi_{+}|-|\varphi_{-}\rangle\langle\varphi_{-}|$.
Obviously, $H_{0}^{a}(t)$ has off-diagonal matrix elements which could cause undesirable nonadiabatic errors. To inhibit the nonadiabatic errors, we make a modification to the original Hamiltonian as
\begin{align}\label{eqb4}
H_m(t)=H_0(t)+H_c(t).
 \end{align}
To find out appropriate $H_c(t)$, firstly, in the adiabatic picture defined by $U(t)$, we have
\begin{align}\label{eqb5}
    H_{U}(t)&=U(t)H_m(t)U^{\dag}(t)+i\dot{U}(t)U^{\dag}(t)\cr
      &=\Omega(t)M_z+\dot{\theta}M_y+U(t)H_c(t)U^{\dag}(t).
\end{align}
 Suppose $H_U(t)$ has the form as $H_U(t)=[g_z(t)+\Omega(t)]M_z+[\dot{\theta}+g_y(t)]M_y+g_x(t)M_x$, where $g_x(t)$, $g_y(t)$, and $g_z(t)$ are time-dependent parameters. According to Eq.~(A5), it is deduced that
\begin{align}\label{eqb6}
 H_{c}(t)&=U^{\dag}(t)[g_x(t)M_x+g_y(t)M_y+g_z(t)M_z]U(t) \cr
 &=[g_z(t)\sin\theta-g_x(t)\cos\theta]|1\rangle\langle3| \cr
 &+ig_y(t)|1\rangle\langle2| \cr
 &+[g_z(t)\cos\theta+g_x(t)\sin\theta]|2\rangle\langle3|+\textmd{H.c.}.
 \end{align}
It is better that $H_c(t)$ possesses the same form as the original Hamiltonian $H_0(t)$, so we set $g_y(t)=0$.

To determine the specific forms of $g_x(t)$ and $g_z(t)$, according to DSBS, we introduce another unitary transformation \cite{BaksicPRL16116}
 \begin{align}\label{eqb8}
    V(t)=\exp[i\beta(t)M_x],
     \end{align}
where $\beta(t)$ is a time-dependent parameter.
Then, in the picture defined by $V(t)$, the Hamiltonian $H_U(t)$ becomes
\begin{align}\label{eqb9}
    &H_V(t)=V(t)H_U(t)V^{\dag}(t)+i\dot{V}(t)V^{\dag}(t)\cr
    &=V(t)[(g_z+\Omega) M_z+\dot{\theta}M_y+g_xM_x]V^{\dag}(t)+i\dot{V}(t)V^{\dag}(t).\cr
     \end{align}
Expressing $V(t)$ as $V(t)=\sum_{k=0,1,2}|\tilde{\xi}_k\rangle\langle\tilde{\xi}_k(t)|$, where $\{|\tilde{\xi}_k\rangle\}$ is a set of time-independent states and $\{|\tilde{\xi}_k(t)\rangle\}$ is a set of time-dependent dressed states in the adiabatic picture, we obtain
\begin{align}
 &|\tilde{\xi}_0(t)\rangle=\frac{1}{\sqrt{2}}[i\sin\beta|\tilde{\xi}_1\rangle+\sqrt{2}\cos\beta|\tilde{\xi}_0\rangle-i\sin\beta|\tilde{\xi}_2\rangle],\cr
 &|\tilde{\xi}_1(t)\rangle=\frac{1}{2}[(1+\cos\beta)|\tilde{\xi}_1\rangle+i\sqrt{2}\sin\beta|\tilde{\xi}_0\rangle+(1-\cos\beta)|\tilde{\xi}_2\rangle],\cr
 &|\tilde{\xi}_2(t)\rangle=\frac{1}{2}[(1-\cos\beta)|\tilde{\xi}_1\rangle-i\sqrt{2}\sin\beta|\tilde{\xi}_0\rangle+(1+\cos\beta)|\tilde{\xi}_2\rangle], \end{align}
 and
 \begin{align}\label{eqb11}
H_V(t)&=[(g_z+\Omega)\cos\beta-\dot{\theta}\sin\beta](|\tilde{\xi}_1\rangle\langle\tilde{\xi}_1|-|\tilde{\xi}_2\rangle\langle\tilde{\xi}_2|)\cr
&+\{[i(g_z+\Omega)\sin\beta+i\dot{\theta}\cos\beta+\dot{\beta}-g_x]|\tilde{\xi}_1\rangle\langle\tilde{\xi}_0|\cr
 &+[i(g_z+\Omega)\sin\beta+i\dot{\theta}\cos\beta-\dot{\beta}+g_x]|\tilde{\xi}_2\rangle\langle\tilde{\xi}_0|\cr
 &+\textmd{H.c.}\}.
 \end{align}

To let $H_V(t)$ be a diagonal matrix, the coefficients of $|\tilde{\xi}_1\rangle\langle\tilde{\xi}_0|$ and $|\tilde{\xi}_2\rangle\langle\tilde{\xi}_0|$ should equal to zero, so we obtain $g_x(t)=\dot{\beta}$ and $g_z(t)=-\Omega-\dot{\theta}\cot\beta$. Substituting $g_x(t)$ and $g_z(t)$ into Eqs.~(A1), (A4), and (A6), we have
\begin{align}\label{eqb13}
H_m(t)&=\Omega_{1m}(t)|1\rangle\langle3|+\Omega_{2m}(t)|2\rangle\langle3|+\textmd{H.c.},
\end{align}
with the modified pulses are
 \begin{align}\label{eqb13}
 \Omega_{1m}(t)&=-\dot{\theta}\sin\theta\cot\beta-\dot{\beta}\cos\theta, \cr
 \Omega_{2m}(t)&=-\dot{\theta}\cos\theta\cot\beta+\dot{\beta}\sin\theta. \end{align}
And, in the original picture, the evolution states can be obtained as $|\xi_k(t)\rangle=U^{\dag}(t)|\tilde{\xi}_k(t)\rangle$. By careful calculation, we obtain
\begin{align}\label{eqb14}
|\xi_0(t)\rangle&=\cos\theta\cos\beta|1\rangle+i\sin\beta|3\rangle-\sin\theta\cos\beta|2\rangle,\cr
|\xi_1(t)\rangle&=\frac{1}{\sqrt{2}}e^{i\varsigma t}[(\sin\theta+i\cos\theta\sin\beta)|1\rangle+i\sqrt{2}\cos\beta|3\rangle \cr
&+(\cos\theta-i\sin\theta\sin\beta)|2\rangle],\cr
|\xi_2(t)\rangle&=\frac{1}{\sqrt{2}}e^{-i\varsigma t}[(\sin\theta-i\cos\theta\sin\beta)|1\rangle-i\sqrt{2}\cos\beta|3\rangle \cr
&+(\cos\theta+i\sin\theta\sin\beta)|2\rangle],
\end{align}
where $\varsigma=-\int_{0}^{t}\frac{\dot{\theta}}{\sin\beta}dt^{\prime}$.

\section{}


In this appendix, we show the specific derivation of the effective Hamiltonian $H_{\textmd{eff}}$ in Eq.~(10). When the initial states are the computational basis states in subspace $\mathcal{Z}_{\pm}$, according to the full Hamiltonian $H_I$ in Eq.~(8), states in
$\mathcal{Z}_{-}=\{|g_{l},g_{l},-\rangle$, $|g_{l},g_{r},-\rangle,$
$|g_{r},g_{l},-\rangle,$
$|g_{r},g_{r},-\rangle\}$ are decoupled from Hamiltonian $H_I$, so they do not evolve. For the states in subspace
 $\mathcal{Z}_{+}=\{|g_{l},g_{l},+\rangle$, $|g_{l},g_{r},+\rangle,$
$|g_{r},g_{l},+\rangle,$
$|g_{r},g_{r},+\rangle\}$, according to the full Hamiltonian $H_I$ in Eq.~(8), these four states evolve independently in four independent subspaces. Therefore, we divide the full evolution space into four subspaces so that $H_I$ is block-diagonal on them. These four independent subspaces are:

(I) $\mathcal{Z}_{+1}=\{|\phi_1\rangle\sim|\phi_{11}\rangle\}$:
\begin{align}
  |\phi_{1}\rangle&=|g_{l},g_{r},+\rangle, ~|\phi_{2}\rangle=|g_{l},g_{r},e\rangle,
  \cr
  |\phi_{3}\rangle&=|g_l,g_{r},g_{r}\rangle|1_r\rangle_{c_3},
  ~|\phi_{4}\rangle=|g_l,g_{r},g_{r}\rangle|1_r\rangle_{f_2},
  \cr
  |\phi_{5}\rangle&=|g_l,g_{r},g_{r}\rangle|1_r\rangle_{c_2},
  ~|\phi_{6}\rangle=|g_l,e,g_{r}\rangle,
  \cr
  |\phi_{7}\rangle&=|g_l,g_{l},g_{r}\rangle|1_l\rangle_{c_2},
  ~|\phi_{8}\rangle=|g_l,g_{l},g_{r}\rangle|1_l\rangle_{f_1},
  \cr
  |\phi_{9}\rangle&=|g_l,g_{l},g_{r}\rangle|1_l\rangle_{c_1},
  ~|\phi_{10}\rangle=|e,g_{l},g_{r}\rangle,
  \cr
  ~~|\phi_{11}\rangle&=|f,g_{l},g_{r}\rangle.
  \end{align}

(II) $\mathcal{Z}_{+2}=\{|\phi_{12}\rangle\sim|\phi_{20}\rangle\}$:
\begin{align}
  |\phi_{12}\rangle&=|g_{r},g_{r},+\rangle, ~ |\phi_{13}\rangle=|g_{r},g_{r},e\rangle,
  \cr
  |\phi_{14}\rangle&=|g_r,g_{r},g_{r}\rangle|1_r\rangle_{c_3},
  ~|\phi_{15}\rangle=|g_r,g_{r},g_{r}\rangle|1_r\rangle_{f_2},
  \cr
  |\phi_{16}\rangle&=|g_r,g_{r},g_{r}\rangle|1_r\rangle_{c_2},
  ~|\phi_{17}\rangle=|g_r,e,g_{r}\rangle,
  \cr
  |\phi_{18}\rangle&=|g_r,g_{l},g_{r}\rangle|1_l\rangle_{c_2},
  ~|\phi_{19}\rangle=|g_r,g_{l},g_{r}\rangle|1_l\rangle_{f_1},
  \cr
  |\phi_{20}\rangle&=|g_r,g_{l},g_{r}\rangle|1_l\rangle_{c_1}.
  \end{align}

(III) $\mathcal{Z}_{+3}=\{|\phi_{21}\rangle\sim|\phi_{25}\rangle\}$:
\begin{align}
  |\phi_{21}\rangle&=|g_{l},g_{l},+\rangle, ~|\phi_{22}\rangle=|g_{l},g_{l},e\rangle,
  \cr
  |\phi_{23}\rangle&=|g_l,g_{l},g_{r}\rangle|1_r\rangle_{c_3},
  ~|\phi_{24}\rangle=|g_l,g_{l},g_{r}\rangle|1_r\rangle_{f_2},
  \cr
  |\phi_{25}\rangle&=|g_l,g_{l},g_{r}\rangle|1_r\rangle_{c_2}.
    \end{align}

(IV) $\mathcal{Z}_{+4}=\{|\phi_{26}\rangle\sim|\phi_{30}\rangle\}$:
\begin{align}
  |\phi_{26}\rangle&=|g_{r},g_{l},+\rangle, ~|\phi_{27}\rangle=|g_{r},g_{l},e\rangle,
  \cr
  |\phi_{28}\rangle&=|g_r,g_{l},g_{r}\rangle|1_r\rangle_{c_3},
  ~|\phi_{29}\rangle=|g_r,g_{l},g_{r}\rangle|1_r\rangle_{f_2},
  \cr
  |\phi_{30}\rangle&=|g_r,g_{l},g_{r}\rangle|1_r\rangle_{c_2}.
    \end{align}

The derivation of $H_{\textmd{eff}}$ is divided into two steps: first, since $H_I$ in Eq.~(8) is a block-diagonal matrix on subspaces $\mathcal{Z}_{+1}\sim\mathcal{Z}_{+4}$, we rewrite $H_I$ as $H_I=\bigoplus\limits_{j=1}^{4}H_{I}^{j}=\bigoplus\limits_{j=1}^{4}(H_{I0}^{j}+H_{al}^{j})$ and express $H_{I0}^{j}$ in the matrix form in the subspace $\mathcal{Z}_{+j}$; next, we perform a unitary transformation $U_{I0}^{j}=e^{i H_{I0}^{j}t}$ on $H_I^{j}$ and obtain
   \begin{align}
  H_{Ie}^{j}=U_{I0}^{j}H_I^{j}U_{I0}^{j\dag}+i\dot{U}_{I0}^{j}U_{I0}^{j\dag}=U_{I0}^{j}H_{al}^{j}U_{I0}^{j\dag}.
    \end{align}

For convenient description, hereafter, we use $\{|\psi_m\rangle\}$ and $\{\eta_m\}~(m=1,2,...,30)$ to represent the eigenstates and eigenvalues of $H_{I0}^{j}$ in different subspaces. For example, the subscript $m=(1,...,11)$ corresponds to the subspace $\mathcal{Z}_{+1}$, the subscript $m=(12,...,20)$ corresponds to the subspace $\mathcal{Z}_{+2}$. Note that $\{\eta_m\}$ usually are the functions of $(\lambda,\nu)$ except some of them are zero.

In the subspace $\mathcal{Z}_{+1}$, by rewriting $H_{I0}^{1}$ in the matrix form, we obtain the eigenvalues
\begin{align}
& \eta_1=0,~\eta_2=0,~\eta_3=0,\cr
&\eta_4=-\sqrt{(\lambda^2+2\upsilon^2-B)/2},~\eta_5=\sqrt{(\lambda^2+2\upsilon^2-B)/2},\cr
&\eta_6=-\sqrt{(3\lambda^2+2\upsilon^2-B)/2},~\eta_7=\sqrt{(3\lambda^2+2\upsilon^2-B)/2},\cr
&\eta_8=-\sqrt{(\lambda^2+2\upsilon^2+B)/2},~\eta_9=\sqrt{(\lambda^2+2\upsilon^2+B)/2},\cr
&\eta_{10}=-\sqrt{(3\lambda^2+2\upsilon^2+B)/2},~\eta_{11}=\sqrt{(3\lambda^2+2\upsilon^2+B)/2},\cr  \end{align}
 with $B=\sqrt{\lambda^4+4\upsilon^4}$. Here, for brevity, we only list the eigenstates with the corresponding eigenvalues equal to $0$, such as eigenstates $|\psi_1\rangle$, $|\psi_2\rangle$, and $|\psi_3\rangle$:
\begin{align}
& |\psi_1\rangle=|\phi_1\rangle, ~~ |\psi_2\rangle=|\phi_{11}\rangle,\cr
&|\psi_3\rangle=N_3(|\phi_{2}\rangle-\frac{\lambda}{\upsilon}|\phi_{4}\rangle+|\phi_{6}\rangle-\frac{\lambda}{\upsilon}|\phi_{8}\rangle+|\phi_{10}\rangle),\cr
\end{align}
  where $N_3=\upsilon/\sqrt{3\upsilon^2+2\lambda^2}$ is the normalization factor of the eigenstate $|\psi_3\rangle$.

Then, we rewrite $H_{I0}^{1}$ and $\{|\phi_n\rangle~(n=1,...,11)\}$ with the eigenstates $\{|\psi_m\rangle~(m=1,2,...,11)\}$ as
\begin{align}
  H_{I0}^{1}&=\sum_{m=1}^{11}\eta_m|\psi_m\rangle\langle\psi_m|=\sum_{m=4}^{11}\eta_m|\psi_m\rangle\langle\psi_m|,\cr
  |\phi_n\rangle&=\sum_{m=1}^{11}\alpha_m^{(n)}|\psi_m\rangle,~n=1,2,...,11,
    \end{align}
  where $\{\alpha_m^{(n)}\}$ are the corresponding expanding coefficients.
  Since $H_{al}^{1}$ can be rewritten as
\begin{align}\label{eqc4}
  H_{al}^{1}=\Omega_3(t)|\phi_1\rangle\langle\phi_2|+\Omega_1(t)|\phi_{11}\rangle\langle\phi_{10}|+\textmd{H.c.},
    \end{align}
thus, according to Eqs.~(B5-B9), we deduce that
\begin{align}
  H_{Ie}^{1}&=U_{I0}^{1}H_{al}^{1}U_{I0}^{1\dag}\cr
  &=\Omega_3(t)[\alpha_3^{(2)}|\psi_1\rangle\langle\psi_3|+\sum_{m=4}^{11}e^{-i\eta_mt}\alpha_m^{(2)}|\psi_1\rangle\langle\psi_m|] \cr
  &+\Omega_1(t)[\alpha_3^{(10)}|\psi_2\rangle\langle\psi_3|+\sum_{m=4}^{11}e^{-i\eta_mt}\alpha_m^{(10)}|\psi_2\rangle\langle\psi_m|] \cr
  &+\textmd{H.c.},
    \end{align}
where $\alpha_3^{(2)}=\alpha_3^{(10)}=N_3$.

In the subspace $\mathcal{Z}_{+2}$, it is not difficult to find that, for Hamiltonian $H_{I0}^{2}$, there is only one eigenstate $|\psi_{12}\rangle=|\phi_{12}\rangle$ with eigenvalue $\eta_{12}=0$. For brevity, the explicit forms of $\eta_m~(m=13,14,...,20)$ are not presented. Following the routine given in the subspace $\mathcal{Z}_{+1}$,
  we rewrite $H_{I0}^{2}$, $H_{al}^{2}$, and $\{|\phi_n\rangle~(n=13,14,...,20)\}$ as
\begin{align}
  H_{I0}^{2}&=\sum_{m=12}^{20}\eta_m|\psi_m\rangle\langle\psi_m|=\sum_{m=13}^{20}\eta_m|\psi_m\rangle\langle\psi_m|,\cr
     H_{al}^{2}&=\Omega_3(t)|\phi_{12}\rangle\langle\phi_{13}|+\textmd{H.c.},\cr
     |\phi_n\rangle&=\sum_{m=12}^{20}\alpha_m^{(n)}|\psi_m\rangle, ~n=12,13,...,20,
  \end{align}
where $\{\alpha_m^{(n)}\}$ are the corresponding expanding coefficients.
Then we deduce
\begin{align}
 & H_{Ie}^{2}=U_{I0}^{2}H_{al}^{2}U_{I0}^{2\dag}\cr
&=\Omega_3(t)\sum_{m=13}^{20}e^{-i\eta_mt}\alpha_m^{(13)}|\psi_{12}\rangle\langle\psi_m|
    +\textmd{H.c.}.
\end{align}

With the similar routine as that in subspace $\mathcal{Z}_{+2}$  , in subspaces $\mathcal{Z}_{+3}$ and $\mathcal{Z}_{+4}$, we will obtain
\begin{align}
 &H_{Ie}^{3}=\Omega_3(t)\sum\limits_{m=22}^{25}e^{-i\eta_mt}\alpha_m^{(22)}|\psi_{21}\rangle\langle\psi_m|+\textmd{H.c.},\cr
 &H_{Ie}^{4}=\Omega_3(t)\sum\limits_{m=27}^{30}e^{-i\eta_mt}\alpha_m^{(27)}|\psi_{26}\rangle\langle\psi_m|+\textmd{H.c.}.\cr
  \end{align}

Since the non-zero $\{\eta_m\}$ are the functions of $\lambda$ and $\upsilon$ as exemplified in Eq.~(B6), therefore, if the condition
$\Omega_{1(3)}(t)\ll (\lambda,\upsilon)$ is fulfilled, the terms which have coefficients $e^{-i\eta_mt}$ in Eqs.~(B10), (B12), and (B13) can be viewed as quick oscillating terms and ignored. Thus, the Hamiltonian $H_{Ie}=\bigoplus\limits_{j=1}^{4}H_{Ie}^{j}$ is simplified as
\begin{align}
    H_{\textmd{eff}}&=N_3\Omega_{3}(t)|\phi_{1}\rangle\langle \psi_{3}|+N_3\Omega_{1}(t)|\phi_{11}\rangle\langle\psi_{3}|+\textmd{H.c.}\cr
    &=\tilde{\Omega}_{3}(t)|\phi_{1}\rangle\langle \psi_{3}|+\tilde{\Omega}_{1}(t)|\phi_{11}\rangle\langle\psi_{3}|+\textmd{H.c.},
    \end{align}
with the effective Rabi frequency
\begin{align}
    \tilde{\Omega}_{1(3)}(t)=N_3\Omega_{1(3)}(t),~N_3=\frac{\upsilon}{\sqrt{3\upsilon^2+2\lambda^2}}.
    \end{align}

\section{}

\begin{figure} [!htb]
 \scalebox{0.16}{\includegraphics{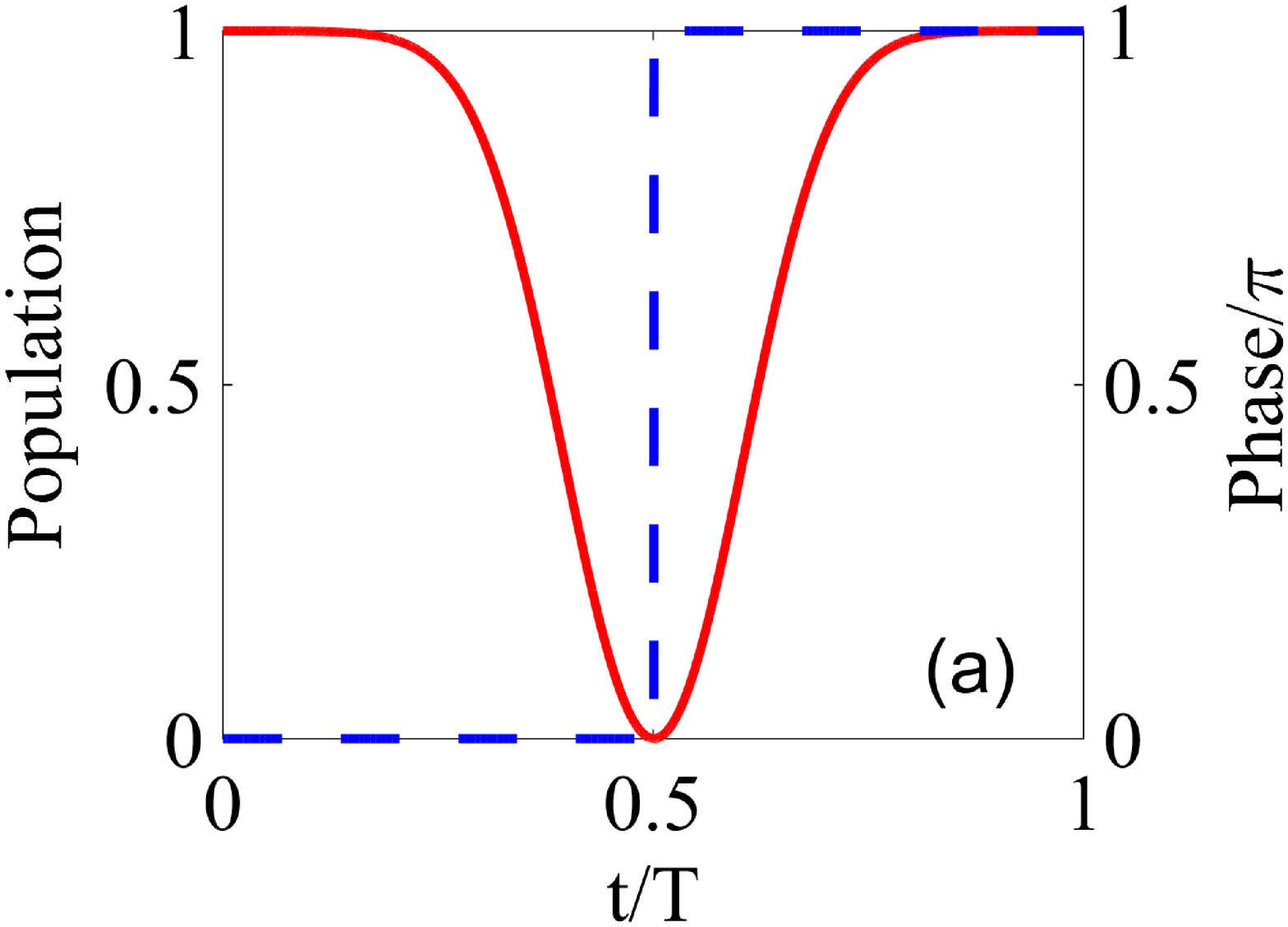}}
 \scalebox{0.16}{\includegraphics{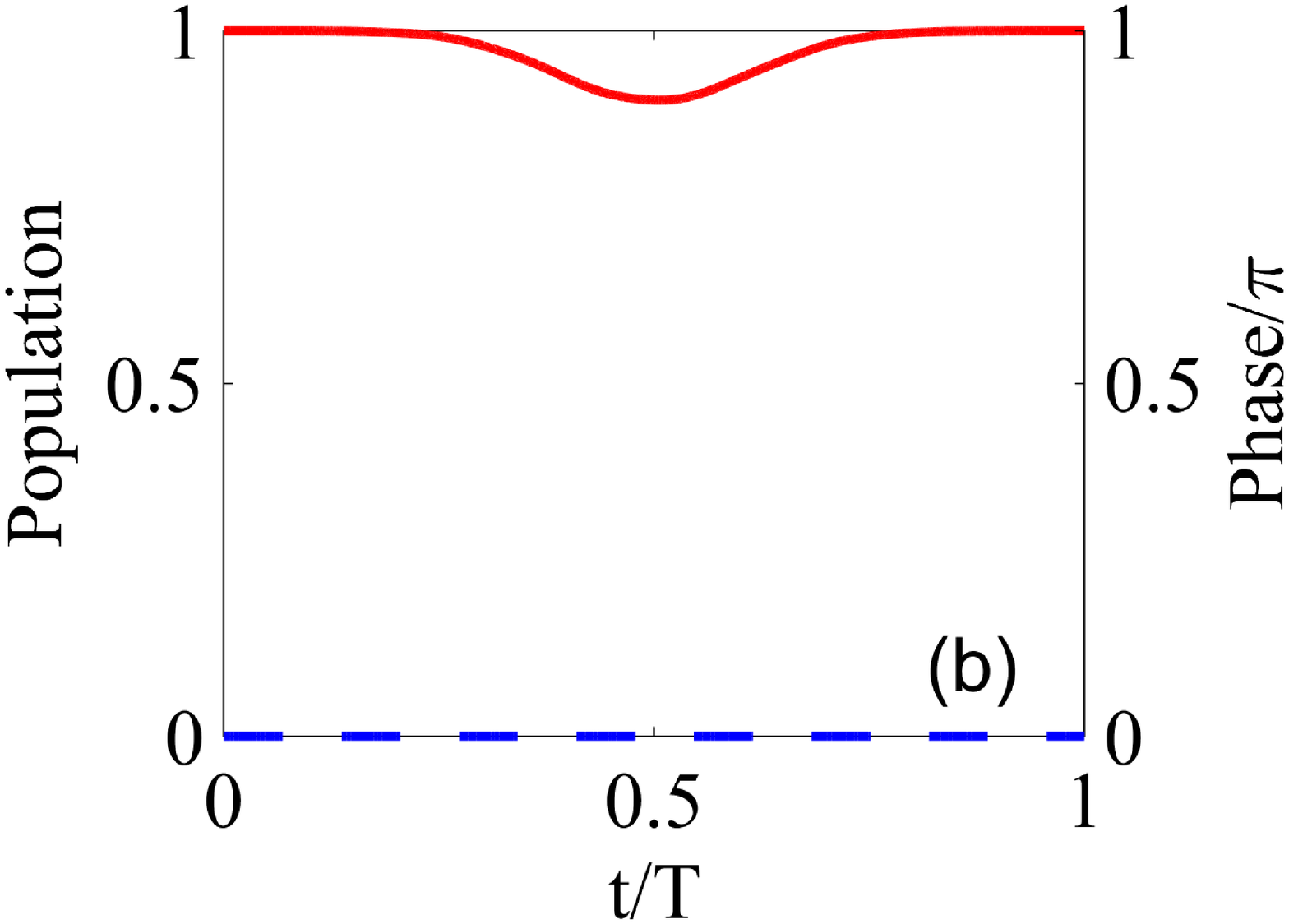}}
 \scalebox{0.16}{\includegraphics{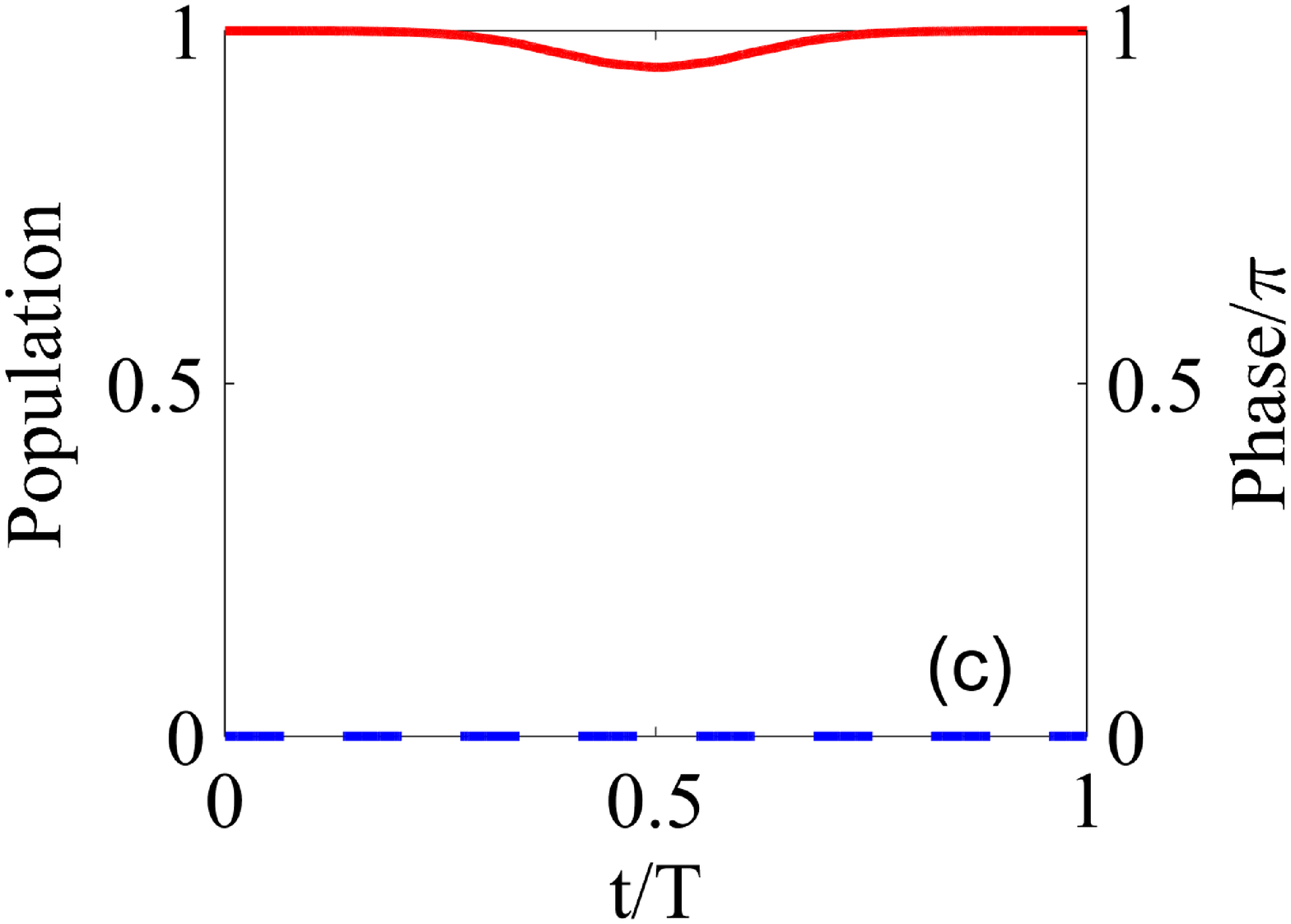}}
 \scalebox{0.16}{\includegraphics{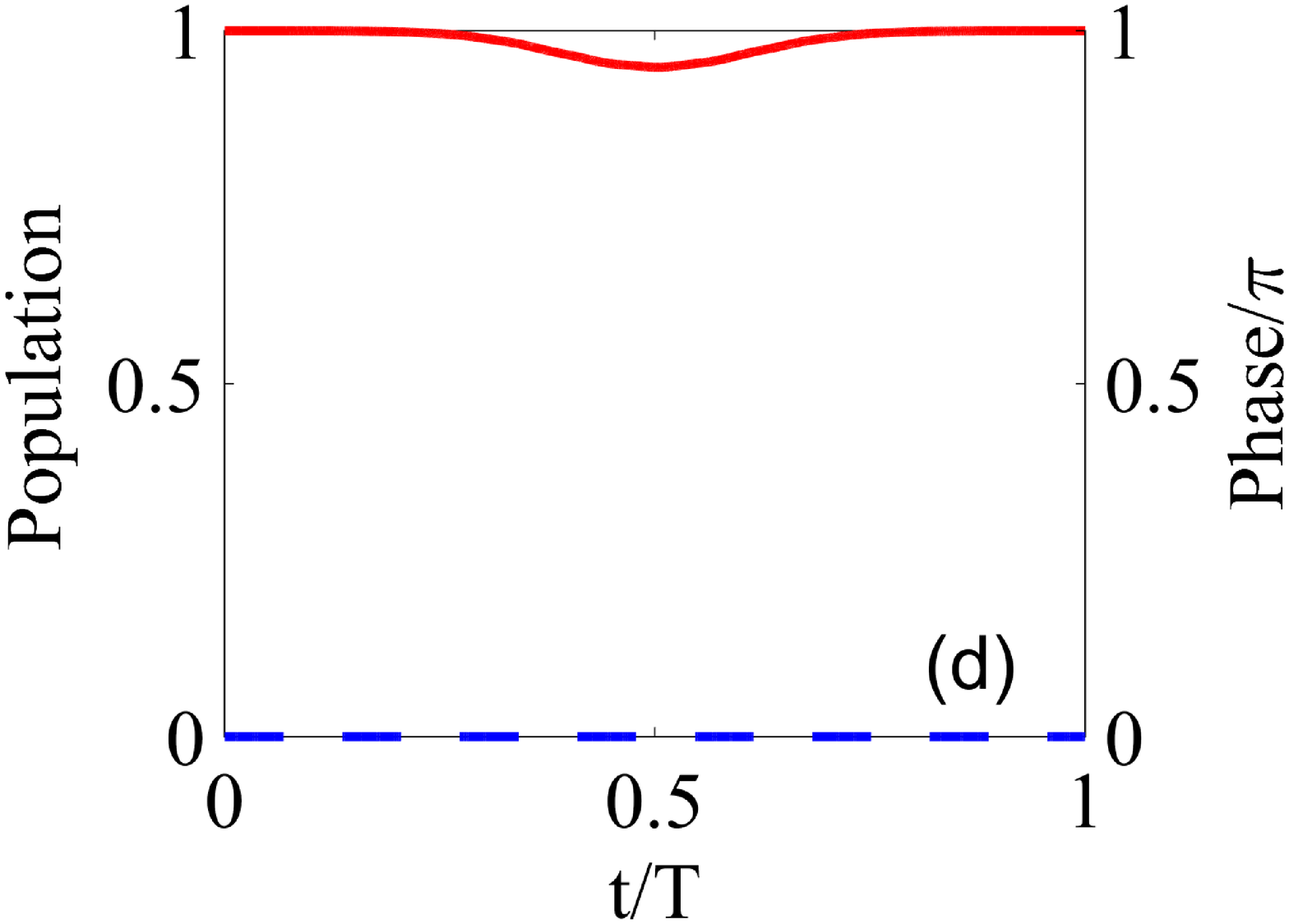}}
  \caption{
         Time-dependent population (solid red) and time-dependent phase (dash blue) of the computational basis state
         (a) $|g_{l},g_{r},+\rangle$;
         (b) $|g_{r},g_{r},+\rangle$;
         (c) $|g_{l},g_{l},+\rangle$;
         (d) $|g_{r},g_{l},+\rangle$. The left y-axis indicates the population, and the right y-axis indicates the phase.
          }
 \label{fig7}
\end{figure}

In this Appendix C, with the pulses $\bar{\Omega}_{1(3)}(t)$ in Eq.~(27), we will numerically show that in the computational basis subspace $\mathcal{Z}_{\pm}=\{|g_{l},g_{l},-\rangle$, $|g_{l},g_{l},+\rangle,$
$|g_{r},g_{l},-\rangle,$
$|g_{r},g_{l},+\rangle,$
$|g_{r},g_{r},-\rangle,$
$|g_{r},g_{r},+\rangle,$
$|g_{l},g_{r},-\rangle,$
$|g_{l},g_{r},+\rangle\}$, the gate operation in Eq.~(14) can be well implemented.

First of all, it is obvious that states $\{|g_{l},g_{l},-\rangle$, $|g_{r},g_{l},-\rangle,$
$|g_{r},g_{r},-\rangle$, and $|g_{l},g_{r},-\rangle\}$ do not evolve since they are decoupled from Hamiltonian $H_I$ in Eq.~(8).
In Fig.~7, based on the pulses $\bar{\Omega}_{1(3)}(t)$ and the Hamiltonian $H_I$ in Eq.~(8), we plot the temporal evolution of the populations and the phases of the relevant computational states in $\mathcal{Z}_{+}$, respectively. As shown in Fig.~7(a), if the initial state is $|g_{l},g_{r},+\rangle$, at the end of evolution, the population of state $|g_{l},g_{r},+\rangle$ also reaches 1 (the solid red line in Fig.~7(a)), but the phase changes $\pi$ (the dash blue line in Fig.~7(a)). However, as shown in Fig.~7(b-d), if the initial states are respectively $|g_{r},g_{r},+\rangle$, $|g_{l},g_{l},+\rangle$, and $|g_{r},g_{l},+\rangle$, in the evolution,
 the populations of theses states
are almost unchanged and the corresponding phases keep invariant. That is, these states hardly evolve in the process.

\section{}

To demonstrate that the present scheme is a shortcut to adiabaticity one, we implement the gate operation $U_g$ in Eq.~(14) with the adiabatic method. The starting point is
the effective Hamiltonian in Eq.~(10)
\begin{align}
    H_{\textmd{eff}}=\tilde{\Omega}_{3}(t)|g_{l},g_{r},+\rangle\langle \psi_{3}|+\tilde{\Omega}_{1}(t)|f,g_{l},g_{r}\rangle\langle\psi_{3}|+\textmd{H.c.}.
    \end{align}
By choosing
\begin{align}
    \tilde{\Omega}_{3}(t)=\tilde{\Omega}_{0}(t)\sin\zeta(t),~\tilde{\Omega}_{1}(t)=\tilde{\Omega}_{0}(t)\cos\zeta(t),
    \end{align}
with $\tilde{\Omega}_{0}(t)=\sqrt{\tilde{\Omega}_{3}(t)^2+\tilde{\Omega}_{1}(t)^2}$ and $\zeta(t)=\arctan(\tilde{\Omega}_{3}(t)/\tilde{\Omega}_{1}(t))$, we can obtain the ``dark state'' (zero-energy eigenstate) of $H_{\textmd{eff}}$ as
\begin{align}
    |\textmd{Dk}(t)\rangle&=\cos\zeta(t)|g_{l},g_{r},+\rangle-\sin\zeta(t)|f,g_{l},g_{r}\rangle.
     \end{align}

To realize the gate operation $U_g$, it should guarantee that the system is initially in the state $|g_{l},g_{r},+\rangle$ and finally in $-|g_{l},g_{r},+\rangle$.
As for standard adiabatic passage, so long as the adiabatic condition $\dot{\zeta}(t)\ll\tilde{\Omega}_{0}(t)$ is satisfied, the system will remain in $|\textmd{Dk}(t)\rangle$ at all the time. Therefore, by evolving $\zeta(t)$ continuously from $0$ to $\pi$, $|\textmd{Dk}(t)\rangle$ will vary from $|g_{l},g_{r},+\rangle$ to $-|g_{l},g_{r},+\rangle$.
We can design pulses to satisfy the boundary condition $\zeta(0)=0,~\zeta(T)=\pi$ as
\begin{align}
    \tilde{\Omega}_{3}(t)&=\tilde{\Omega}\sin^2\frac{\pi t}{T},\cr
    \tilde{\Omega}_{1}(t)&=\tilde{\Omega}\sin\frac{\pi t}{T}\cos\frac{\pi t}{T},
     \end{align}
where $\tilde{\Omega}$ is the pulse amplitude. Then, to compare the present scheme with the adiabatic one, we plot Fig.~8 to show the average fidelity $\bar{F}_e$ versus $t/T$ with different methods. For better comparison, the amplitudes of the pulses used in adiabatic passage are approximately equal to that of the pulses designed by the dressed states method, and the coupling strength is selected as $\lambda=150/T$ for both of the methods. As shown in Fig.~8, the average fidelity $\bar{F}_e$ of the present scheme reaches 1 at $t=T$ (see the solid blue line in Fig.~8), while for adiabatic passage, $\bar{F}_e$ reaches 1 at $t\approx5T$  (see the dotted red line in Fig.~8). Therefore, the present paper proposes a shortcut scheme to obtain the target.
\begin{figure}[!htb]
 \scalebox{0.23}{\includegraphics{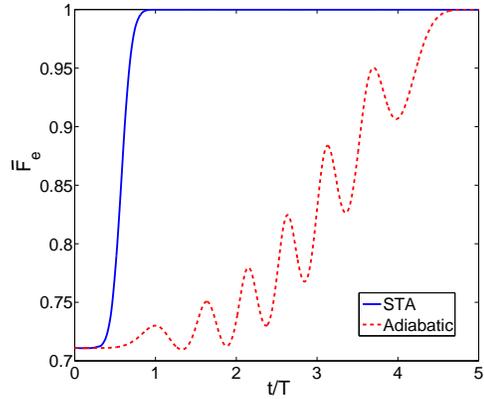}}
   \caption{
         The average fidelity $\bar{F}_{e}$ of the gate operation $U_g$ versus $t/T$ with different methods: solid blue line, using the dressed states STA method with pulses $\bar{\Omega}_{1(3)}$
         ; dotted red line, using the adiabatic passage with pulses $\tilde{\Omega}_{1(3)}(t)$,  $\tilde{\Omega}=30/T$ and $\lambda=150/T$.
          }
 \label{fig8}
\end{figure}

\end{document}